\documentclass[%
 preprint,
superscriptaddress,
 amsmath,amssymb,
pre,
]{revtex4-1}
\usepackage{graphicx}
\usepackage{dcolumn}
\usepackage{bm}
\usepackage[mathlines]{lineno}
\usepackage{color}
\usepackage[colorlinks=true,allcolors=blue,breaklinks=true]{hyperref}
\usepackage{physics}

\begin{document}

\title{Localized growth drives spongy mesophyll morphogenesis}

\author{John D. Treado}
\email{john.treado@yale.edu}
\altaffiliation[Present address: ]{Max Planck Institute for the Physics of Complex Systems, Dresden, Germany}
\affiliation{Department of Mechanical Engineering \& Materials Science and Integrated Graduate Program in Physical and Engineering Biology, Yale University, New Haven, CT, USA, 06520}

\author{Adam B. Roddy}
\affiliation{Institute of Environment, Department of Biological Sciences, Florida International University, Miami, FL, USA, 33199
}

\author{Guillaume Th\'{e}roux-Rancourt}
\affiliation{University of Natural Resources and Life Sciences, Vienna, Department of Integrative Biology and Biodiversity Research, Institute of Botany, 1180 Vienna, Austria}

\author{Liyong Zhang}
\affiliation{Department of Biology, College of Arts and Science, University of Saskatchewan, Saskatoon, Canada, S7N 5E2}

\author{Chris Ambrose}
\affiliation{Department of Biology, College of Arts and Science, University of Saskatchewan, Saskatoon, Canada, S7N 5E2}

\author{Craig Brodersen}
\affiliation{School of the Environment, Yale University, New Haven, CT, USA, 06520}

\author{Mark D. Shattuck}
\affiliation{Department of Physics and Benjamin Levich Institute, City College of New York, New York, NY, USA, 10031}

\author{Corey S. O'Hern}
\email{corey.ohern@yale.edu}
\affiliation{Department of Mechanical Engineering \& Materials Science and Integrated Graduate Program in Physical and Engineering Biology, Yale University, New Haven, CT, USA, 06520}
\affiliation{Department of Physics and Department of Applied Physics, Yale University, New Haven, CT, USA, 06520}

\begin{abstract}
The spongy mesophyll is a complex, porous tissue found in plant leaves that enables carbon capture and provides mechanical stability. Unlike many other biological tissues, which remain confluent throughout development, the spongy mesophyll must develop from an initially confluent tissue into a tortuous network of cells with a large proportion of intercellular airspace. How the airspace in the spongy mesophyll develops while the cells remain mechanically stable remains unknown. Here, we used computer simulations of deformable particles to develop a purely mechanical model for the development of the spongy mesophyll tissue. By stipulating that (1) cell perimeter grows only near voids, (2) cells both form and break adhesive bonds, and (3) the tissue pressure remains constant, the computational model was able to recapitulate the developmental trajectory of the microstructure of the spongy mesophyll observed in \emph{Arabidopsis thaliana} leaves. Robust generation of pore space in the spongy mesophyll requires a balance of cell growth, adhesion, stiffness and tissue pressure to ensure cell networks remain both porous yet mechanically robust. The success of this mechanical model of tissue growth and porosity evolution suggests that simple physical principles can coordinate and drive the development of complex plant tissues like the spongy mesophyll.
\end{abstract}

\maketitle

\section{Introduction}\label{sec:intro}
\begin{figure}[!t]
    \centering
    \includegraphics[width=0.8\textwidth]{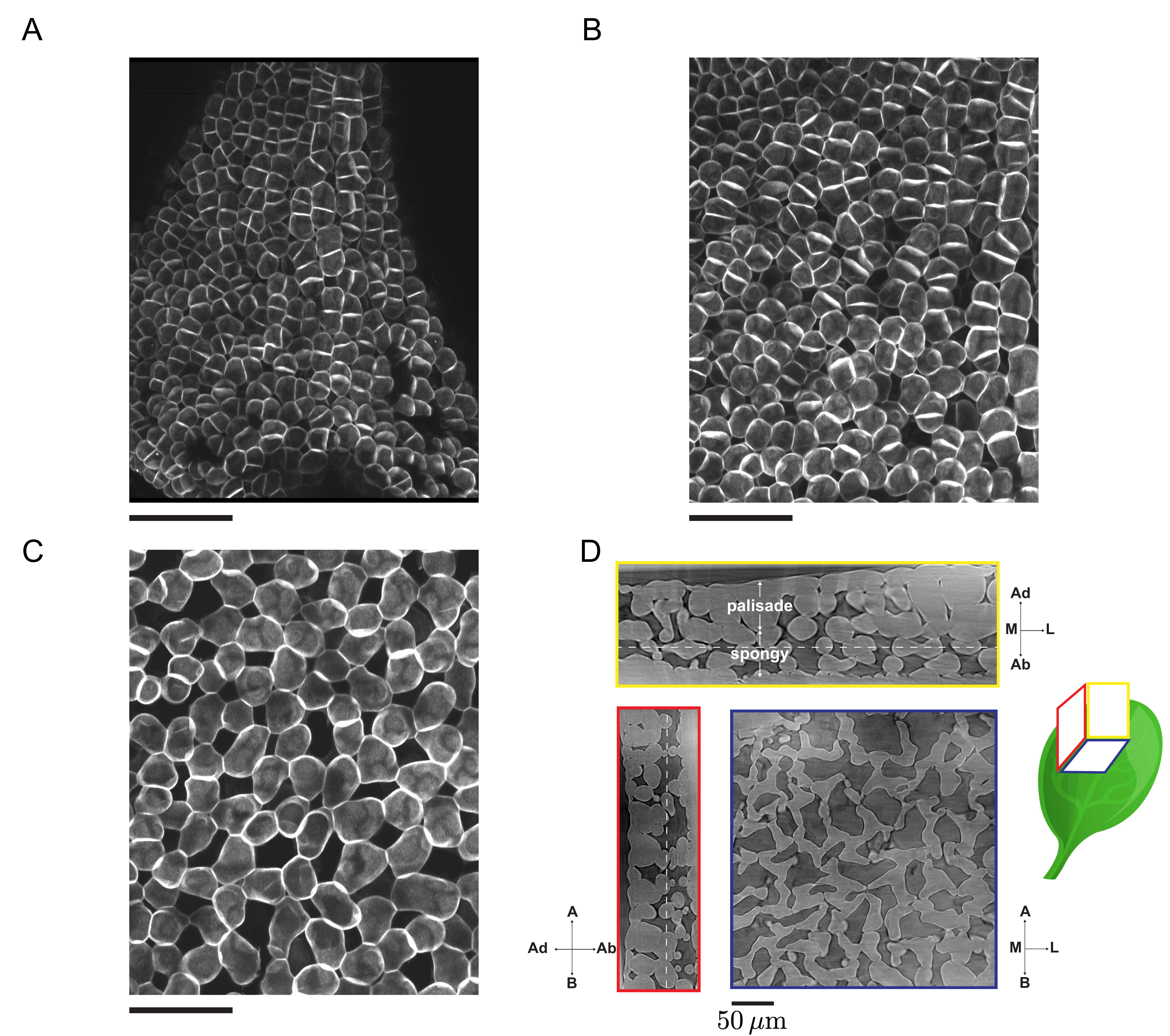}
    \caption{Confocal microscopy images of the developing spongy mesophyll in \emph{Arabidopsis thaliana} taken at (A) 0, (B) 24, and (C) 72 hours of development. (See methods for details.) The black scale bar in each frame represents 50 $\mu$m. (D) Mesophyll tissue observed in a microcomputed tomography (microCT) scan of a mature \emph{Arabdidopsis} leaf. The leaf has three orthogonal axes, the basal-apical (BA), medial-lateral (ML), and adaxial-abaxial (AdAb) axes. Leaf images are in the three planes orthogonal to these axes, i.e. the transverse (yellow), longitudinal (red), and paradermal (purple) planes, respectively. The paradermal image is taken at the dashed white lines drawn on the other two planar slices.}
    \label{fig:exp}
\end{figure}
Morphogenesis, or the emergence of structure during biological development, requires the careful coordination of cell growth and motility across entire tissues. Plant tissues, in particular, require significant coordination because plant cells are not motile~\citep{physio:AveryAmJBot1993}. During development, plant cells undergo often dramatic changes in their size, shape, and number of neighbors during development~\citep{shape:GreenJCellBio1965,physio:SeagoAnnBot2005,shape:IvakovFrontInPlantSci2013,plants:SapalaELife2018,plants:AntreichAdvSci2019,plantevo:RoddyIJPS2020}. For example, leaf epidermal pavement cells expand by many orders of magnitude during leaf development and take on a diversity of shapes, depending on global factors, such as organ growth anisotropy, and local factors, such as the mechanical stresses imposed by their immediate neighbors~\citep{physio:HamantScience2008,physio:SampathkumarCurrBio2014,plants:SapalaELife2018,plants:VofelyNewPhyto2019}. While there have been significant advances in characterizing and modeling tissue morphogenesis in plants, these efforts have been focused almost entirely on tissues composed of confluent cells~\citep{plants:RoederNatRevMCB2011,cellmodel:BoudonPLOSCompBio2015}. However, one of the most important tissues in plants, the mesophyll tissue inside leaves where photosynthesis occurs, is a porous assembly of both cells and intercellular air space. After entering the leaf through the stomata on the leaf surface, CO$_2$ must diffuse through the airspace and into the cells, where it is converted into sugar ~\citep{plants:LundgrenNatComm2019,physio:TherouxRancourtProcRoySocB2021}. While the upper layer of mesophyll, the palisade mesophyll, is composed of densely packed, elongated cells with the same orientation, the lower, spongy mesophyll is highly porous and composed of cells with many different shapes and orientations ~\citep{plants:ScottBotGaz1948,plants:BorsukNewPhyto2022,plants:TherouxRancourtNewPhyt2020}, leading to variation in carbon assimilation~\citep{plants:TerashimaJPlantResearch2001,plants:EvansPCE2003,plants:LehmeierThePlantJ2017}. Yet, the spongy mesophyll cells begin development densely packed as convex polyhedra with nearly spherical cell shapes~\citep{physio:PsarasNewPhyto1995,dev:PanterisNewPhyto2005,physio:ZhangPlantCell2021}.  The cell mechanics and cell-cell interactions that control the formation of pore space in the spongy mesophyll are currently unknown ~\citep{plant:WhitewoodsPLOSBIO2021}.

Recent experimental advances in three-dimensional imaging have shed light on the development of complex cell shapes in both young and mature leaves~\citep{physio:WuytsPlantMethods2010,physio:KalveJExpBot2014,physio:ZhangPlantCell2021}. Early in development, spongy mesophyll tissue is densely packed with each cell taking on a convex, polygonal shape in paradermal cross-section (Fig.~\ref{fig:exp}). However, as the tissue grows, airspace begins to form between cells, and the tissue transitions from nearly confluent to porous. At maturity, cells in the spongy mesophyll form tortuous, quasi-two-dimensional (quasi-2D) porous networks~\citep{plants:BorsukNewPhyto2022,physio:TherouxRancourtProcRoySocB2021,plants:HarwoodPCE2021}. The pore space forms due to cell expansion and the breaking of cell-cell contacts, not programmed cell death~\citep{physio:PsarasNewPhyto1995}. While cell adhesion~\citep{physio:DaherFrontInPlantSci2015}, tissue pressure~\citep{plant:ZimmermannPlanta1980}, and cell shape change~\citep{physio:HamantScience2008,physio:EngCurrOpPlantBio2018,physio:ZhangPlantCell2021} are known to be important for morphogenesis of the plant epidermis, it is unclear if or how they contribute to development of the spongy mesophyll. Computational modeling of tissue morphogenesis can provide answers to these open questions, and to our knowledge no model has ever been developed for plant tissues that can vary in porosity along a developmental trajectory.  

Here, we (1) develop computer simulations of deformable polygons with shape degrees of freedom~\citep{jamming:BoromandPRL2018,jamming:TreadoPRM2021} in two dimensions to recapitulate the developmental trajectory of the spongy mesophyll, (2) compare these simulations to empirical images of spongy mesophyll development in  the model plant species \emph{Arabidopsis thaliana} (L.) Heynh, and (3) test the sensitivity of our developmental model to variation in cell growth, cell-cell adhesion, and bulk tissue pressure. This approach reveals three key features that are necessary to capture the development of the quasi-2D spongy mesophyll microstructure observed in \emph{Arabidopsis thaliana} leaves. First, while cell areal growth is constant throughout the tissue and during development, growth and remodelling of cell wall must be localized to cell boundaries \emph{not} in contact with other cells, i.e. exposed to the intercellular airspace~\citep{physio:ZhangPlantCell2021}. Second, cells must balance cell wall bending rigidity with cell-cell adhesive strength to develop networks with evenly spaced pores. If cell walls are too stiff, or, counter-intuitively, if adhesion is too weak and contacts break too frequently, then the networks either collapse or fracture due to the formation of a single, large pore. Third, the pressure inside the simulation boundary must be positive throughout the developmental trajectory. While cells expand during development due to positive turgor pressure~\citep{plants:NiklasUChicago2012}, remodelling of cell wall near voids leads to localized growth that can cooperatively push the tissue boundary outward. That is, the change in the simulation domain size is not an independent variable in our simulations but instead arises from differences in the growth rates of cell perimeter and cell area. This computational model therefore demonstrates that a porous tissue as complex as the spongy mesophyll can be assembled by following a simple set of mechanical rules. 

\section{Results}\label{sec:results}
\subsection{Adhesive, growing deformable polygons at constant pressure generate porous cell networks}

\begin{figure}[!t]
    \centering
    \includegraphics[width=0.8\textwidth]{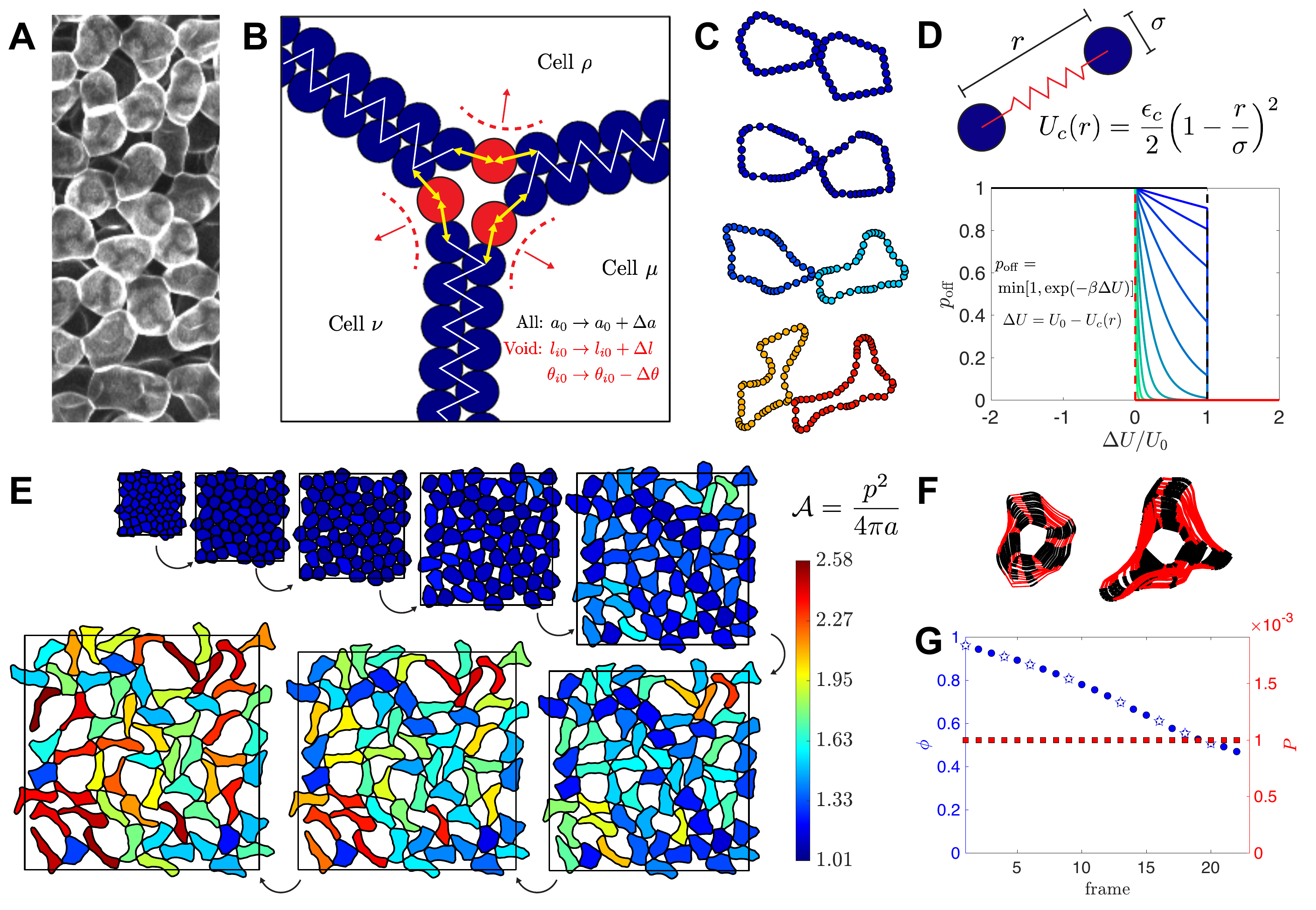}
    \caption{Summary of model for spongy mesophyll development in two dimensions. (A) Close-up of cell shape during spongy mesophyll development, taken from Fig.~\ref{fig:exp} (C). (B) Close-up of DP model cell boundaries, highlighting the three main rules of growth. (See main text for details.) (C) DP model cells changing shape during the simulation of spongy mesophyll development, from top to bottom. Vertices of each DP model cell are drawn as disks with radius $\sigma$, as defined in panel D. Disk color is given by the DP cell shape parameter $\mathcal{A} = p^2/4\pi a$, and disk size is rescaled to be the same in each row. (D) Summary of how intracellular bonds break as a function of the inverse effective temperature $\beta$ and breaking energy $U_0$. (See main text for details.) (E) Schematic of developmental trajectory from DP model simulations: $64$ cells growing in a square boundary (black line) with periodic boundary conditions. Cells and the boundary area are drawn to scale at each step, and cell color represents the cell shape parameter. (F) The boundaries of two cells from (E) drawn to scale. Black boundary segments indicate vertices in contact with other cells, whereas red boundary segments indicate vertices in contact with void. (G) Comparison of packing fraction $\phi$ and pressure $P$ during the simulation in (E). Stars indicate specific frames from (E). The simulation in (E)-(G) used $\epsilon_b = 0.4$, $\beta = 100$, $h = 2.4$. $\Delta a = 0.5$, $\lambda = 5$, $c_L = 0.5$, $c_B = 4$, $\theta_{0,{\rm min}} = -\pi/10$, and $P_0 = 10^{-3}$. See Table~\ref{tab:params} for parameter descriptions.}\label{fig:model}
    \raggedright
\end{figure}

To model the development of the spongy mesophyll in two spatial dimensions (2D), we employ numerical simulations of deformable polygons (the DP model, see~\citep{jamming:BoromandPRL2018,jamming:TreadoPRM2021}) that can change shape in response to stress as well as interact with each other via repulsive and attractive forces. Our simulations contain a collection of $N$ polygons, each with $n$ vertices. The dynamics of the polygons, which we will refer to as cells, are governed by forces due to changes in cell shape and forces due to cell-cell interactions. The potential energy  $U_{\mu, {\rm shape}}$ for changes in cell shape for cell $\mu$ is defined as
\begin{equation}\label{eq:ushape}
    U_{\mu, {\rm shape}} = \frac{\epsilon_a}{2}\qty(\frac{a_\mu}{a_{0\mu}} - 1)^2 + \sum_{i=1}^{n_\mu} \qty[\frac{\epsilon_l}{2}\qty(\frac{l_{i\mu}}{l_{0i\mu}} - 1)^2 + \frac{\epsilon_b}{2}\qty(\theta_{i\mu} - \theta_{0i\mu})^2],
\end{equation}
where $\epsilon_a$, $\epsilon_l$, and $\epsilon_b$ control the amount of energy required to have cell areas $a$, perimeter segment lengths $l$, and local curvatures $\theta$ that deviate from their preferred values ($a_0$, $l_0$, and $\theta_0$, respectively). Each segment length is labeled $l_{i\mu}$ and bending angle $\theta_{i\mu}$, where $i = 1,\ldots,n_\mu$. A graphical representation of these quantities, as well as derivations of the forces due to each term in Eq.~\ref{eq:ushape}, are provided in section S$1$ in the Supplementary Information (SI). The total potential energy $U$ is the shape potential energy plus the potential energy contribution from cell-cell interactions:
\begin{equation}\label{eq:utotal}
    U = \sum_{\mu = 1}^N \qty[U_{\mu, {\rm shape}} + \sum_{\nu > \mu} \sum_{i=1}^{n_\mu}\sum_{j=1}^{n_\nu} \frac{\epsilon_{ij\mu\nu}}{2}\qty(1 - \frac{r_{ij\mu\nu}}{\sigma_{ij\mu\nu}})^2].
\end{equation}
Each vertex $i$ on cell $\mu$ is treated as a soft disk with diameter $\sigma_{i\mu}$. The parameter $\epsilon_{ij\mu\nu}$ controls the energy required for vertex $i$ on cell $\mu$ to change its distance $r_{ij\mu\nu}$ to vertex $j$ on cell $\mu$. $\sigma_{ij\mu\nu} = (\sigma_{i\mu} + \sigma_{j\nu})/2$ is the distance at which vertex $i$ is in contact with vertex $j$. If $r_{ij\mu\nu} < \sigma_{ij\mu\nu}$, the vertices overlap. Two vertices will experience an attractive interaction when $r_{ij\mu\nu} > \sigma_{ij\mu\nu}$ only if $\epsilon_{ij\mu\nu} > 0$. Because mesophyll cells adhere to their neighbors~\citep{physio:LionettiPhytochem2015, physio:DaherFrontInPlantSci2015}, we model cell-cell adhesion as harmonic springs and set $\epsilon_{ij\mu\nu} > 0$ when two vertices are "bonded". Because contact-breaking helps form void space in the developing mesophyll~\citep{physio:ZhangPlantCell2021}, cells are allowed to break adhesive bonds, as described in section S$2$ in the SI. Briefly, a bond is broken either when the bond energy $U_c$ exceeds $U_0$ or when the bond breaks stochastically with probability $p_{\rm off} = \min[1,\exp(-\beta\Delta U)]$, where $\Delta U = U_0 - U_c$ and $\beta$ is an inverse effective temperature scale. Thus, the probability of bond-breaking is dependent on $\beta$ (Fig.~\ref{fig:model}D). In these simulations, energies and lengths are measured in units of $\epsilon_a$ and the averaged square root of the cell area, i.e. $\rho = N^{-1} \sum_\mu \sqrt{a_{0\mu}}$. All vertices have unit mass, and time is measured in units of $\rho \epsilon_a^{-1/2}$. 

\begin{table}[t]
    \centering
    \small
    \begin{tabular}{|c|c|c|c|c|}
        \hline
        Symbol & Meaning & Units & Value\\
        \hline
        $\epsilon_a$ & area elasticity & energy unit & $1$\\
        $\epsilon_l$ & perimeter elasticity & $\epsilon_a$ & $1$\\
        $\epsilon_c$ & contact elasticity & $\epsilon_a$ & $1$\\
        $\epsilon_b$ & bending elasticity & $\epsilon_a$ & $10^{-2} - 1$ ($0.4$)\\
        $\mathcal{A}_{\rm max}$ & shape parameter limit & $n\pi^{-1}\tan(n^{-1}\pi)$ & $3$\\
        $\beta$ (section S$2$ in SI) & inverse effective adhesion temperature & $\epsilon_a^{-1}$ & $20 - 1000$ ($100$)\\
        $r^*$ (section S$2$ in SI) & relative bond fracture length & $\sigma$ & $2 - 4$ ($2.4$)\\
        $\Delta a$ (section S$3$ in SI) & cell area growth increment & $\rho^2$ & $10^{-4} - 0.75$ ($0.5$)\\
        $\lambda$ (section S$3$ in SI) & void perimeter growth scale & $n^{-1}\rho^{-2} p\Delta a $ & $1 - 10$ ($5$)\\
        $c_L$ (section S$3$ in SI) & perimeter growth regularization & $1$ & $0 - 1$ ($0.5$) \\
        $c_B$ (section S$3$ in SI) & scale of curvature change & $n^{-1}\rho^{-3} p\Delta a $ & $1 - 10$ ($4$)\\
        $\theta_{0,{\rm min}}$ (section S$3$ in SI) & minimum bending angle & rads. & $-\pi/4 - 0$ ($-\pi/10$)\\
        $P_0$ & boundary pressure & $\epsilon_a \rho^{-2}$ & $10^{-7} - 0.1$ ($10^{-3}$)\\
        \hline
    \end{tabular}
    \caption{Table of model parameters. Appendices (that provide a more complete description) are listed next to parameters that are not defined fully in the main text. Units of $1$ signify dimensionless quantities. In the final column, numbers in parentheses indicate the values of the parameters in the simulation in Figs.~\ref{fig:model} and~\ref{fig:comp}. We show in Figs.~\ref{fig:dadl} and~\ref{fig:kbbe} that the values selected for $\epsilon_b$, $\beta$, $\Delta a$, and $\lambda$ are nearly optimal for recapitulating the developmental trajectory of spongy mesophyll in \emph{Arabidopsis thaliana}.}
    \label{tab:params}
\end{table}

Our simulations begin with densely packed cells that reflect the nearly confluent, nascent spongy mesophyll tissue, and they grow to form porous cell networks (Fig.~\ref{fig:model}E)~\citep{plants:SiftonBotRev1945,physio:PsarasNewPhyto1995,dev:PanterisNewPhyto2005,physio:ZhangPlantCell2021}. To form a dense initial state, we isotropically compress dilute configurations of cells by small compression and potential energy minimization steps, similar to previous work on model jammed materials~\citep{jamming:GaoPRE2006,jamming:BoromandPRL2018}, to a near-confluent state with $< 3$\% of the simulation domain occupied by void space. Once the initial near-confluent state is generated, adhesive bonds are formed between all overlapping vertices, and all spontaneous curvatures $\theta_{0}$ at each vertex are set to their instantaneous $\theta$ values. This point defines the initial state with characteristics reflective of the earliest images of mesophyll development~\citep{physio:PsarasNewPhyto1995,plants:RhizopoulouAnnBot2003}. 

The full details of our model for plant cell growth and shape-change are provided in section S$3$ in the SI, but are briefly summarized here. At the beginning of a growth step, the \emph{preferred} area $a_0$ of each cell is increased by an increment $\Delta a$, i.e.
\begin{equation}\label{eq:del_a0}
    a_0 \to a_0 + \Delta a.
\end{equation}
Because of the area energy term in Eq.~\eqref{eq:utotal}, instances where $a_0$ exceeds the instantaneous cell area $a$ induces a positive internal pressure and drives cells to grow in size. We also actively change the \emph{preferred} perimeter length segments $l_0$ each growth step by an increment $\Delta l$, i.e.
\begin{equation}\label{eq:del_l0}
    l_0 \to l_0 + \Delta l.
\end{equation}
Because cell boundaries grow more quickly where they are adjacent to void space than where they are in contact with other cells~\citep{physio:ZhangPlantCell2021}, we grow only the preferred length $l_0$ of the perimeter length segments (Fig.~\ref{fig:model}B, yellow arrows) whose vertices lack contacts with vertices of other DP cells (Fig.~\ref{fig:model}B, red disk). $\Delta l$ depends on both a dimensionless scale parameter $\lambda$, which determines how fast the void-facing cell perimeter grows relative to the cell area, and $c_L$, which regulates how fast a given void segment grows compared to the other void segments (see Table~\ref{tab:params}, and see further details in section S$3$ in the SI). We set $\Delta l = 0$ when the particle shape parameter exceeds a threshold $\mathcal{A}_{\rm max}$, which we set to $3$, close to the maximum value of individual spongy mesophyll cells ($\sim 2.9$)~\citep{plants:TherouxRancourtNewPhyt2020,plants:HarwoodNewPhyto2020}. In addition, the void-facing spontaneous curvature $\theta_0$ is updated according to
\begin{equation}
    \theta_0 \to \theta_0 - \Delta \theta,
\end{equation}
where $\Delta \theta > 0$. $\theta_0$ \emph{decreases} nears voids at each growth step, which drives cell shapes towards increasingly lobed structures, reminiscent of cell shapes in porous network configurations~\citep{plants:SiftonBotRev1945,plants:ScottBotGaz1948,physio:PsarasNewPhyto1995,plants:BorsukNewPhyto2022} (dashed red lines in Fig.~\ref{fig:model}B). We set $\Delta\theta = 0$ when the curvature becomes less than a target minimum value $\theta_{0,{\rm min}}$ (section S$3$ in SI).

This model aims to recapitulate a vital feature of plant tissue development, that positive turgor pressure coupled with cell wall remodelling drives both plant cell growth and plant cell shape change~\citep{plants:NiklasUChicago2012}. We induce positive turgor pressure in each cell by growing $a_0$, and remodel our simulated cell walls by changing $l_0$ and $\theta_0$. Our assumption that the preferred lengths of local cell wall segments $l_0$ only \emph{increase} is reminiscent of the classic Lockhardt model of plant cell growth~\citep{growth:GorielySpringer2017}, where cell walls plastically deform and only grow when a threshold pressure is met.

While these growth rules induce changes to the \emph{preferred} geometric features of each cell, we need to couple growth to relaxational processes to allow stresses to balance. Therefore, after each growth step, we relax the total \emph{enthalpy} $H = U + P_0A$ for simulation domain area $A$ at constant pressure $P_0 > 0$ and potential energy $U$. The presence of a positive pressure $P_0 > 0$ at the simulation boundary mimics the compressive effect that other tissues, such as the palisade mesophyll and epidermis, would have on the expanding spongy mesophyll. Using a modified version of the FIRE energy minimization algorithm (section S$3$ in the SI), a local minimum of the potential energy is found at a fixed pressure $P_0$ after preferred areas, void-facing preferred perimeters, and void-facing spontaneous curvatures are updated. By finding a minimum of the enthalpy $H$ instead of the potential energy $U$, the simulation domain area $A$ will fluctuate to maintain the pressure at $P_0$. Maintaining a positive tissue pressure throughout a plant's life cycle is important in both individual cells and developing tissues~\citep{plant:ZimmermannPlanta1980,plants:ZoniaCellBiochemBiophys2006,plants:KroegerPLOSONE2011,plants:NiklasUChicago2012}. 

A typical developmental trajectory for a simulation of $64$ DP cells shows the changes in domain size, cell area, and cell shape that occur (Fig.~\ref{fig:model}E). This model results in the simulation domain growing in size faster than the cell areas because the packing fraction
\begin{equation}
    \phi = \frac{\sum_\mu a_\mu}{A},
\end{equation}
is a monotonically decreasing function during the simulation (Fig.~\ref{fig:model}G). Note that the porosity, or fraction of air space in the simulation, is simply $1-\phi$. This model reveals an important quality of growing tissues: localizing growth near voids and imposing a constant pressure boundary condition drives dense packings of deformable, adhesive cells toward porous states. It is important to note that the rate of packing fraction decrease is not stipulated in the model, such as via constant strain on the domain area that is set to be larger than the cell growth rate. Rather, the decrease in $\phi$ (Fig.~\ref{fig:model}E) is due entirely to the growth of void-facing perimeter segments. Also, the pressure remains constant throughout the simulation at $P = 10^{-3}$, matching the input pressure $P_0$ (Fig.~\ref{fig:model}G). The supplementary movie shows the progression of the simulation shown in Fig.~\ref{fig:model}E both with periodic image cells shown in a stationary domain (``tissue frame") and in a domain that grows in size as would be seen from an external observer (``lab frame"), as well as the packing fraction $\phi$ and pressure $P$ over the course of the simulation. 

\subsection{Simulations match cell shapes in experiments}
\begin{figure}[!t]
    \centering
    \includegraphics[width=0.8\textwidth]{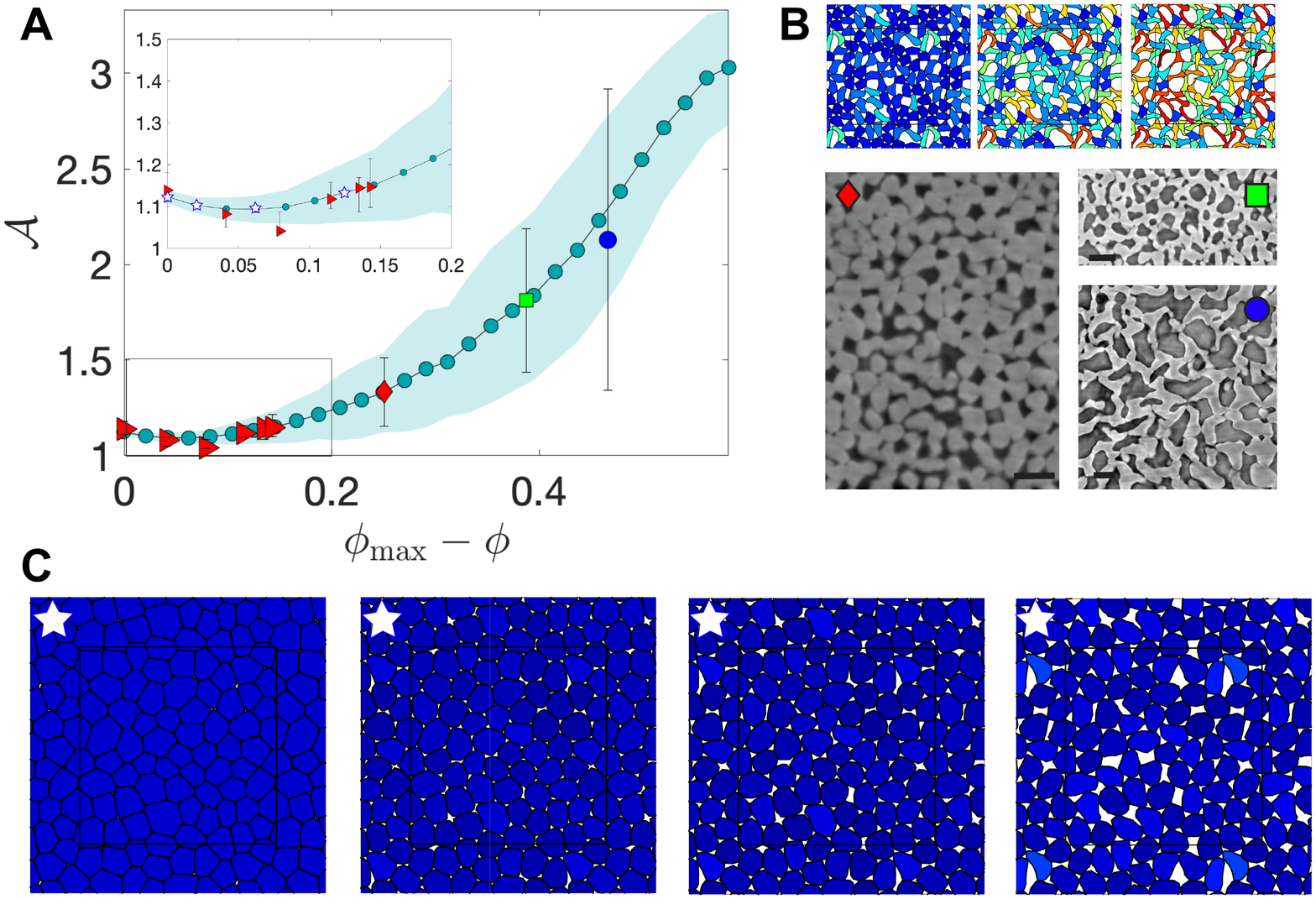}
    \caption{Comparison between simulations of the DP model of spongy mesophyll development and experimental observations. (A) Cell shape parameter $\mathcal{A}$ plotted against $\phi_{\rm max} - \phi$ for a typical developmental trajectory from simulations (green circles, same as Fig.~\ref{fig:model}) and experimental characterization of 2D cell shape projections in \textit{Arabidopsis} at various packing fractions. $\phi_{\rm max} = 1$ for experiments, and $0.975$ for simulations. Red triangles are from 2D projections reported in ~\citep{physio:ZhangPlantCell2021}, and the other markers represent $\mathcal{A}$ of cell projections from segmented microCT scans. The inset is a zoom-in of $\mathcal{A}$ versus $\phi_{\rm max} - \phi$ to highlight the comparison to early developmental stages. (B) Simulation snapshots (top row) and microCT scans (bottom row, greyscale) shown at similar packing fractions. Each scan is labelled by the corresponding plot marker in (A), and simulations decrease in $\phi$ from left to right. Scale bars in each microCT scan are 50 $\mu$m. (C) Snapshots of simulations at low $\phi$ (corresponding to stars in the inset to (A)), decreasing in packing fraction from left to right. In (B) and (C), cell color indicates $\mathcal{A}$ using the colorbar in Fig.~\ref{fig:model}E, and the simulation boundary is periodic and drawn as a black square.}
    \label{fig:comp}
\end{figure}

The decrease in packing fraction of the tissue while maintaining a constant, positive pressure indicates that the DP model captures essential, qualitative features of spongy mesophyll development and produces simulated cells similar in shape to those in developing \emph{Arabidopsis thaliana} leaf spongy mesophyll~\citep{physio:ZhangPlantCell2021} (Fig.~\ref{fig:comp}). In addition to capturing qualitative aspects of spongy mesophyll development, the model also recapitulates quantitative features of spongy mesophyll development. To quantify and compare cell shapes, we measure the cell \emph{asphericity} or shape parameter, 
\begin{equation}
    \mathcal{A} = \frac{p^2}{4\pi a},
\end{equation}
where $p$ and $a$ are a given cell's (or cell projection's) perimeter and area, respectively. $\mathcal{A} = 1$ for circles and $\mathcal{A} > 1$ for all other two-dimensional shapes, e.g. $\mathcal{A} = 1.1$ for a regular hexagon and $\mathcal{A} = 1.65$ for an equilateral triangle. The average shape parameter of cell projections from \emph{Arabidopsis thaliana} leaves at varying developmental stages corresponded closely to the average $\mathcal{A}$ of cells in the developmental simulation shown in Fig.~\ref{fig:model}, changing from dense tissues early in development (Ref.~\citep{physio:ZhangPlantCell2021}) to more porous tissues as the tissue matures (Fig.~\ref{fig:comp}A).  We quantified the agreement between the results from simulations and experiments using the root-mean-square deviation (RMSD) of the cell shape parameter, 
\begin{equation}
    \delta\mathcal{A} = \sqrt{\frac{\sum_{k=1}^m \qty(\mathcal{A}^{\rm sim}_k - \mathcal{A}_k^{\rm exp})^2}{m}},
\end{equation}
where the $k$th experimental data point $\mathcal{A}_k^{\rm exp}$ can be compared to the simulation data point $\mathcal{A}_k^{\rm sim}$ with the most similar $\phi$, and we average over $m = 9$ experimental time points. Over the full range of packing fraction for which we have experimental data, there was less than $4\%$ deviation in $\mathcal{A}$ between simulations and experiments.

\begin{figure}[!t]
    \centering
    \includegraphics[width=0.95\textwidth]{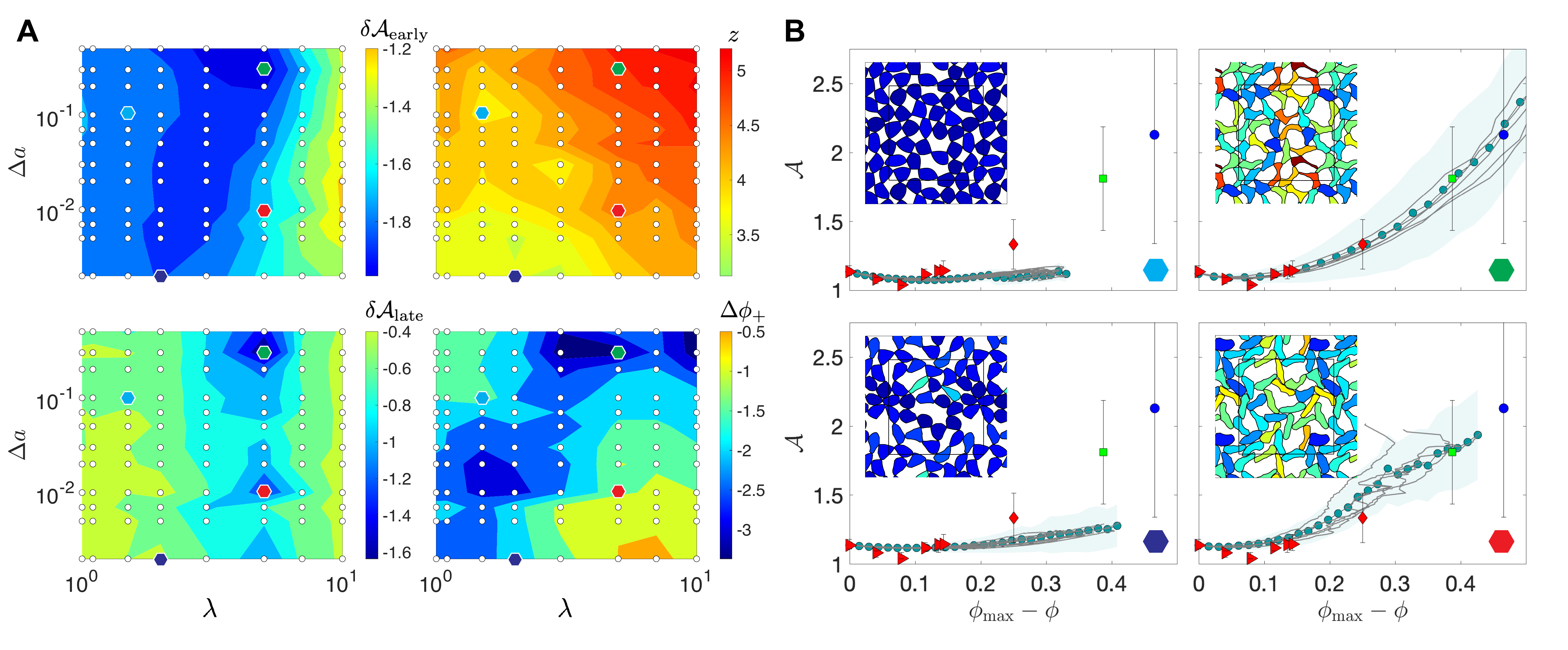}
    \caption{Effect of changing the areal growth $\Delta a$ and void perimeter growth scale $\lambda$ on the shape parameter, packing fraction, and coordination number obtained in simulations. (A) Heatmaps of the RMSD of $\mathcal{A}$ from early developmental stages ($\mathcal{A}_{\rm early}$, top-left), late developmental stages ($\mathcal{A}_{\rm late}$, bottom-left), cumulative compaction ($\Delta \phi_+$, bottom-right) and mean coordination number closest to $\phi = 0.5$ ($z$, top-right). Color represents $\log_{10}$ of each quantity, the values corresponding to each color are indicated by the colorbars, and white circles represent values used to generate the heatmaps. In each heatmap, hexagons placed at $(\lambda, \Delta a) = (1.5,0.1)$ (cyan), $(2,0.002)$ (purple), $(5,0.01)$ (red), and $(5,0.5)$ (green) highlight exemplary simulations shown in panel B. (B) Plots of $\mathcal{A}$ averaged over cells (gray lines) for each simulation, and an ensemble average (green circles) with shading denoting the standard deviation. Hexagons in the lower right corners indicate the particular ensemble parameters (i.e. $\Delta a$ and $\lambda$) highlighted in panel A. Each sub-panel shows a snapshot from the simulation closest to $\phi_{\rm max} - \phi = 0.465$, i.e. that of the latest stage experimental data point (blue circle in Fig.~\ref{fig:comp}A-B. In each inset, cell color denotes $\mathcal{A}$ as defined in the colorbar of Fig.~\ref{fig:model}E. Other parameters (defined in Table~\ref{tab:params}) for these simulations are $\epsilon_b=0.4$, $\beta=100$, $h = 2.4$, $c_L = 0.5$, $c_B = 4$, $\theta_{0,{\rm min}} = -3\pi/20$, and $P_0 = 10^{-6}$. }
    \label{fig:dadl}
\end{figure}

The DP model was also able to recapitulate and explain the transient decline in the shape parameter from $\mathcal{A} \approx 1.15$ to $\mathcal{A} \approx 1.05$ during the initial stages of \emph{Arabidopsis thaliana} leaf development, as well as the subsequent increase in $\mathcal{A}$ after this initial decline (Fig.~\ref{fig:comp}). Cells are initially convex and confluent, and therefore they must have a shape parameter of $\mathcal{A} \approx 1.15$~\citep{jamming:BoromandPRL2018}. Because cell areas always grow but cells grow perimeter only near void space, cell perimeters do not grow when void space is rare. Thus, the cell perimeter does not grow in early stages, and $\mathcal{A}$ must decrease. However, as cells now prefer to be circular, pore space \emph{must} open up as circular shapes cannot tile a plane. Once void space has sufficiently increased, cells can now grow their perimeters near voids, and eventually $\mathcal{A}$ begins to increase again. Our model captures and explains the emergence of non-monotonic behavior of the shape parameter $\mathcal{A}$ during development despite monotonic decreases in the packing fraction $\phi$.

The model succeeded in capturing key dynamics of the developing spongy mesophyll tissue. First, it recovered the non-monotonic behavior of $\mathcal{A}$ in early developmental stages, suggesting that our assumptions regarding spongy mesophyll cell growth and interactions are valid.  Second, the model accurately captured how adhesive, individual cells drive the opening of pore spaces. Third, the model successfully reproduced average shapes of mature cells at low packing fraction, down to $\phi \approx 0.5$ (Fig.~\ref{fig:comp}). For this comparison, we calculated the average $\mathcal{A}$ of individual spongy mesophyll cells that had been segmented from X-ray microcomputed tomographic (microCT) scans of mature \emph{Arabidopsis thaliana} leaves.  Together, these observations indicate that the features of the DP cell model are sufficient to reproduce several important quantitative and qualitative aspects of mesophyll development from the earliest stages of development to maturity.

\subsection{Sensitivity of network formation to parameter variation}

The agreement between the structural properties of spongy mesophyll tissue and of our model results indicates that localized growth of cell perimeters near voids, cell adhesion, and constant boundary pressure can indeed drive the self-assembly of the spongy mesophyll. We further tested the robustness of morphogenesis by characterizing how the structural properties of the tissue change with variation in the model parameters. In particular, we were interested in understanding the range of parameters that can recapitulate \emph{Arabidopsis thaliana} development and whether the packing fraction consistently decreases for any set of model parameters. Addressing these questions is important not only for understanding the development of spongy mesophyll in \emph{Arabidopsis thaliana}, but also for understanding whether the development of the spongy mesophyll in other species can be described by this model~\citep{plants:BorsukNewPhyto2022}.

To probe the sensitivity of the model, we varied two parameters at a time while setting all other model parameters to be similar to those used in Figs.~\ref{fig:model} and~\ref{fig:comp}. Comparisons of the average shape parameter $\mathcal{A}$ of simulations to experimental observations were split into early (i.e. $\delta\mathcal{A}_{\rm early}$) and late (i.e. $\delta\mathcal{A}_{\rm late}$) developmental stages, with the first $m = 6$ experimental data points (red triangles in Fig.~\ref{fig:comp}) used to calculate $\delta\mathcal{A}_{\rm early}$ and the last $m = 3$ data points (other symbols in Fig.~\ref{fig:comp}) used to calculate $\delta\mathcal{A}_{\rm late}$. In addition to comparing $\mathcal{A}$ of simulated and experimental tissues, we also calculated the cumulative compaction $\Delta\phi_+$ as the sum of all positive changes in $\phi$ between growth steps and the average number of cell-cell contacts, $z$, at the simulation snapshots closest to $\phi = 0.5$. While $\Delta\phi_+$ indicates whether simulated tissues expanded or contracted, $z$ is the coordination number that represents the average number of contacts per cell.

First, we tested how varying $\Delta a$, the rate of areal growth, and $\lambda$, the rate of void-facing perimeter growth, influence mesophyll network formation (Fig.~\ref{fig:dadl}). Interestingly, regions of parameter space that minimized RMSD of $\mathcal{A}$ were not always continuous, suggesting that these model parameters can undergo large stochastic jumps with minimal effects on phenotypic outcomes. RMSD of $\delta\mathcal{A}_{\rm early}$ and $\delta\mathcal{A}_{\rm late}$ were minimized in the region where $\Delta a \sim 0.5$ and $\lambda \sim 5$ (green hexagon), with significant deviations in cell shapes of \emph{Arabidopsis thaliana} mesophyll in almost all other regions of parameter space.  While any value of $\Delta a$ led to low $\delta\mathcal{A}$ when $\lambda = 5$, low values of $\Delta a$ led to developmental trajectories with increasing packing fraction rather than decreasing packing fraction, as indicated by $\Delta\phi_+$.  Simulations with $\Delta a = 0.01$ and $\lambda = 5$ (red hexagon) show significant tissue compaction rather than expansion (Fig.~\ref{fig:dadl}B). Compaction likely results from buckling events during development, when a change in the cell shape parameter can make the system mechanically unstable.  Simulations with lower $\Delta a$ had fewer cell-cell contacts on average (Fig.~\ref{fig:dadl}A), suggesting they were less stable. The reduction in $z$ when $\Delta a < 0.01$ could result from there being more simulation steps, which causes an increased probability of adhesive bond breaking. Thus, having a sufficient number of adhesive bonds and allowing sufficient growth to occur along void-facing perimeters are vital for the proper development of spongy mesophyll tissue in \emph{Arabidopsis thaliana}.

\begin{figure}[!t]
    \centering
    \includegraphics[width=0.95\textwidth]{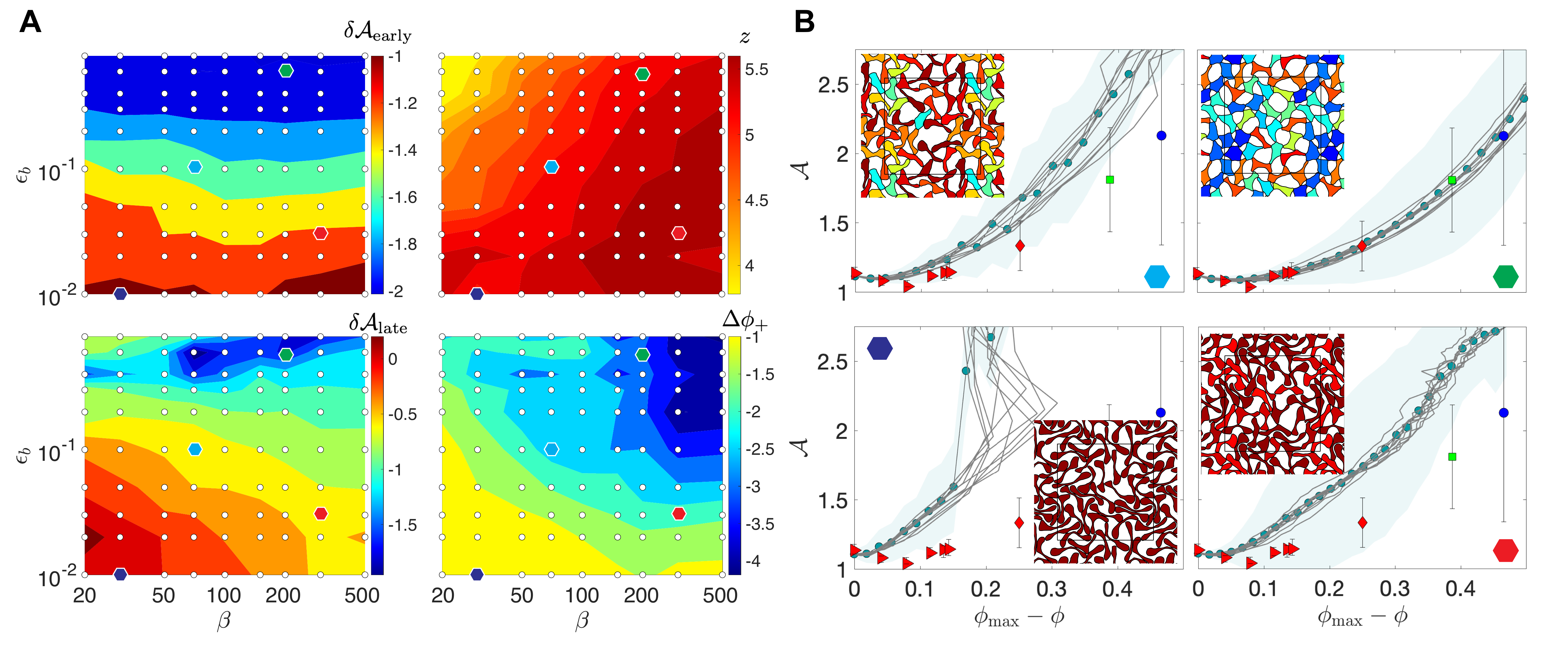}
    \caption{Effect of changing the inverse adhesion temperature $\beta$ and bending energy $\epsilon_b$ on the shape parameter, packing fraction, and coordination number obtained in simulations. (A) Heatmaps of the shape parameter RMSD from early developmental stages ($\mathcal{A}_{\rm early}$, top-left) and late developmental stages ($\mathcal{A}_{\rm late}$, bottom-left), cumulative compaction ($\Delta \phi_+$, bottom-right), and mean coordination closest to $\phi = 0.5$ ($z$, top-right). The color represents $\log_{10}$ of each quantity, and the values corresponding to each color are shown in the colorbars. Hexagons placed at $(\beta, \epsilon_b) = (70,0.1)$ (cyan), $(30,0.01)$ (purple), $(300,0.03)$ (red), and $(200,0.6)$ (green) highlight four exemplary simulations, shown in (B). (B) Plots of $\mathcal{A}$ versus $\phi_{\rm max} - \phi_{\rm max}$ (with snapshots from the simulations in the insets) for the highlighted points in the $(\epsilon_b, \beta)$ parameter space. In each inset, cell color denotes $\mathcal{A}$ as defined in the colorbar of Fig.~\ref{fig:model}E. Other parameters (defined in Table~\ref{tab:params}) for these simulations are $r^* = 2.4\sigma$, $\Delta a=0.5$, $\lambda=5$, $c_L = 0.5$, $c_B = 4$, $\theta_{0,{\rm min}} = -3\pi/20$, and $P_0 = 10^{-7}$.}
    \label{fig:kbbe}
\end{figure}

Second, we tested how varying the strength of the bending energy along cell boundaries $\epsilon_b$ (see Eq.~\eqref{eq:ushape}) and the inverse adhesion temperature $\beta$ influence mesophyll network formation (Fig.~\ref{fig:kbbe}). The inverse adhesion temperature $\beta$ defines the conditions when bonds break, with $\beta \to 0$ meaning that any stretched bond will break and $\beta \to \infty$ meaning that only bonds exceeding the breaking energy $U_0$ will break (Fig.~\ref{fig:model}D). This sensitivity testing revealed that $\beta$ and $\epsilon_b$ must be sufficiently large to generate the correct self-assembly of \emph{Arabidopsis thaliana}-like cell networks (Fig.~\ref{fig:kbbe}A). While the heatmap of $\delta A_{\rm early}$ suggests that $\beta$ has little effect on \emph{Arabidopsis thaliana}-like mesophyll development at low $\phi$, only a small fraction of simulations with large $\epsilon_b$ and large $\beta$ generated correct cell shapes toward the end of development, indicating how developmental progression causes parameter space contraction. Furthermore, decreasing $\beta$, thus losing adhesive contacts, causes tissue compaction during development.  This compaction likely occurs due to buckling events during development (e.g. systems labels with cyan and purple hexagons in Fig.~\ref{fig:kbbe}B). 

These sensitivity tests indicate that generating stable networks without significant buckling events requires both rigid cells \emph{and} adhesive contacts (Figs.~\ref{fig:dadl} and~\ref{fig:kbbe}). As void perimeter growth drives tissue boundary growth in our model, stronger cell perimeter length segments provide rigid scaffolds that can more effectively open pore spaces. However, larger values of $\beta$, which reduce the likelihood of breaking bonds between cells, also lead to the consistent formation of stable, porous cell networks. Successful development of the spongy mesophyll tissue requires that stressed adhesive contacts collectively push the boundary \emph{outward}. That stressed, adhesive bonds can drive expansion seems counter-intuitive. However, pre-stressed bonds often rigidify floppy networks~\citep{pebble:JacobsPRE1996,ellipse:DonevPRE2007,jamming:TreadoPRM2021}. Developing cell networks, therefore, may locally rigidify some regions of the tissue with stretched bonds to push outward rather than inward on the tissue boundary.
\begin{figure}[!t]
    \centering
    \includegraphics[width=0.95\textwidth]{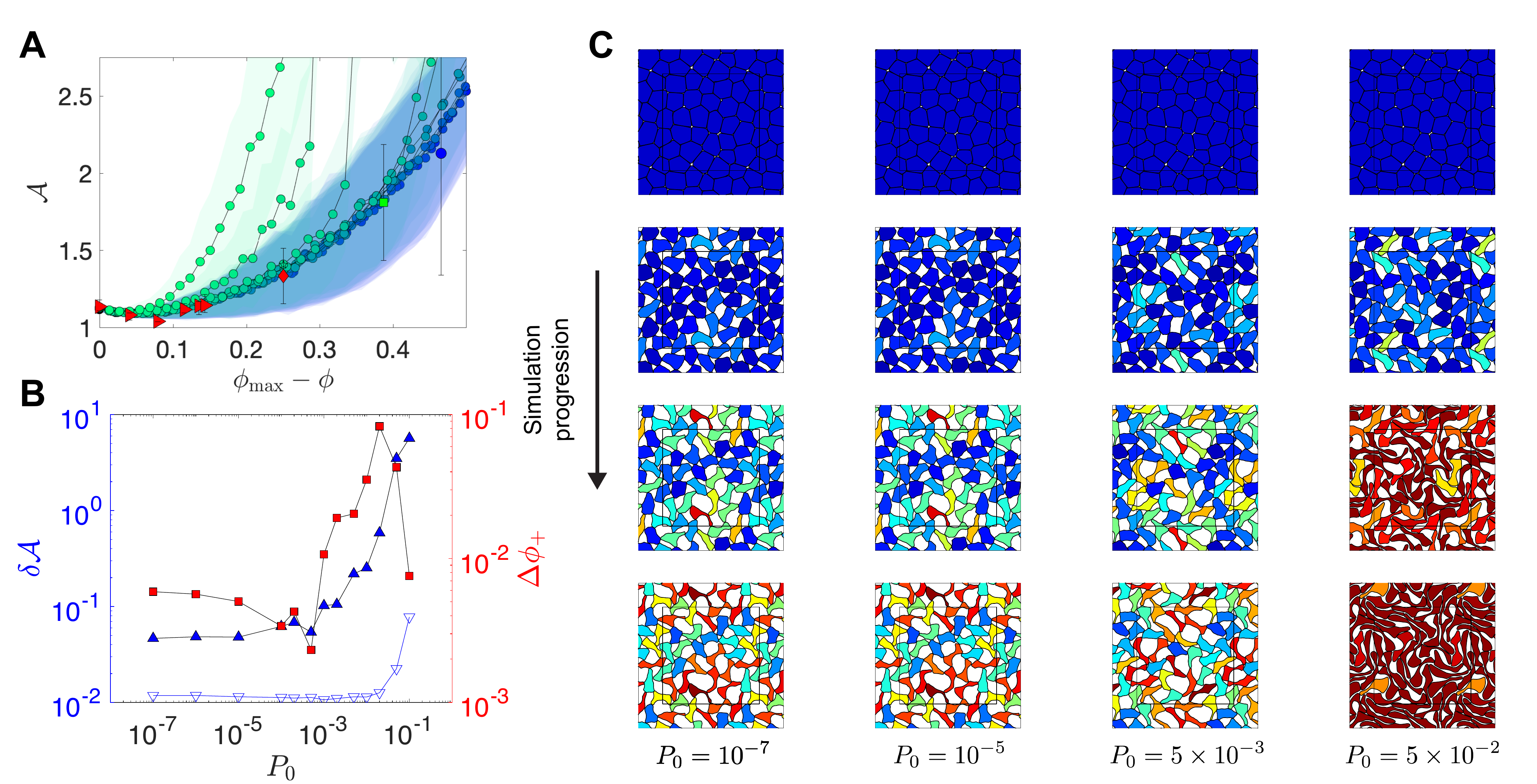}
    \caption{Effect of changing the boundary pressure $P_0$ on the structural properties during development in simulations. (A) Mean cell shape parameter $\mathcal{A}$ for simulations of $32$ cells for $P_0$ between $10^{-7}$ and $10^{-1}$, sorted from low (blue) to high (green). (B) The shape parameter RMSD ($\delta \mathcal{A}$, left axis, blue triangles) and cumulative compaction ($\Delta\phi_+$, right axis, red squares) for simulations at different $P_0$. Blue downward triangles denote $\delta \mathcal{A}_{\rm early}$ and blue upward triangles denote $\delta \mathcal{A}_{\rm late}$. (C) Snapshots of simulations at various pressures drawn using periodic boundary conditions, where the cell color denotes $\mathcal{A}$ with the colorbar in Fig.~\ref{fig:model}. Snapshots are taken at the first frame (row $1$, $\phi = 0.975$), and frames with $\phi$ values closest to those for the three microCT scans in Fig.~\ref{fig:comp}B, i.e. $\phi = 0.75$, $0.6$, and $0.52$. Other parameters for these simulations are $\epsilon_b = 0.4$, $\beta = 100$, $h = 2.4$, $\Delta a = 0.5$, $\lambda = 5$, $c_L = 0.5$, $c_B = 4$, and $\theta_{0,{\rm min}} = -3\pi/20$. }
    \label{fig:pressure}
\end{figure}

Self-organization of mechanical stresses is vital to the correct opening of pore space during spongy mesophyll development. To test the effect of bulk mechanical stress on this process, we varied the external boundary pressure $P_0$ and measured its effect on the structural properties of the tissue during development. Increasing pressure either negligibly affected the simulation results or caused compaction of the tissue (Fig.~\ref{fig:pressure}). When the imposed boundary pressure $P_0 \lesssim 10^{-3}$, the structural properties during development were weakly affected by changes in $P_0$. For example, tissue microstructures were virtually identical when $P_0 = 10^{-7}$ and $10^{-5}$). However, increasing the external pressure above $P_0 = 10^{-3}$ caused the average cell shape parameter to increase more rapidly than that observed in \emph{Arabidopsis thaliana} and led to dramatic increases in the cumulative compaction $\Delta\phi_+$.  For example, when $P_0 = 5\times10^{-2}$, the system compacts, rather than expands in later stages, with cells taking on extremely non-spherical shapes (Fig.~\ref{fig:pressure}). Under such high pressures, growing perimeter segments are unable to push out on the boundary and instead buckle into the void space. The resultant tissue does not expand as much, and can even begin to compact, highlighting the importance of pressure regulation in spongy mesophyll development. Because cell perimeters continue to grow, cells assume biologically unrealistic shapes with extremely large values of $\mathcal{A}$.

The physiological units of the boundary pressure can be obtained by approximating the energy cost for volume changes of hypothetical three-dimensional cells as
\begin{equation}
    U_{\rm 3D} = \frac{\epsilon}{2}\qty(\frac{V}{V_0} - 1)^2,
\end{equation}
where $V_0$ is a cell's preferred volume and $\epsilon$ sets the energy cost for volume fluctuations. If we assume that $U_{\rm 3D}$ is the dominant contribution to isotropic compression, each cell's bulk modulus is 
\begin{equation}
    B \approx -V_0\pdv[2]{U_{\rm 3D}}{V} = \frac{\epsilon}{V_0}.
\end{equation}
If we extend our two-dimensional cells into the third dimension, we can approximate $V_0 \sim a_0^{3/2}$ and $\epsilon = \epsilon_a$, that is, the energy of area deviations.  Since $\epsilon_a$ and $\sqrt{a_0}$ are the units for energy and length scales, the cell bulk modulus is then the unit of pressure in our simulations. If we assume that plant cells have an instantaneous bulk modulus set by the bulk modulus of water, the unit of pressure in our simulations should be $\approx 1$ GPa. The internal turgor pressure in spongy mesophyll cells has been observed to be $\sim 10^{-1}$ MPa~\citep{plant:ZimmermannPlanta1980,plants:FrenschPlanta1988,plants:NonamiPlanta1989,plants:ThurmerProtoplasma1999,physio:RoddyNewPhyto2019}, which is consistent with the scale of cell turgor pressure in other tissues~\citep{plants:KutscheraJPlantPhys2007,plants:NiklasUChicago2012}, and corresponds to $10^{-4}$ in the units of our simulations. Thus, our observation that development is affected by the boundary pressure when $P_0 \gtrsim 10^{-3}$ suggests that the \emph{tissue} pressure must not exceed the \emph{cell} turgor pressure by a factor of $10$ or more. While we know of no measurements of spongy mesophyll boundary pressure, this analysis provide a reasonable upper bound on the tissue pressure. 

\section{Discussion}\label{sec:discussion}
In this paper, we have presented a model of spongy mesophyll development using simulations of deformable polygons that illuminates key cell-level processes driving morphogenesis. This model is, to our knowledge, the first quantitative developmental model of the leaf mesophyll. Our model employs localized growth of cell perimeter to expand the tissue boundary, which is held at constant pressure, such that the internal tissue becomes more porous. This growth process allows tissues to be driven to highly porous states of packing fraction $\phi = 0.5$ or less, while remaining mechanically stable. The success of the model in recapitulating the developmental trajectory of the mean cell shape parameter $\mathcal{A}$ of \emph{Arabidopsis thaliana} from the earliest stages of development to maturity provides insights into the dynamics of mesophyll cell and tissue development. Furthermore, by varying model parameters, we found that maintaining adhesive contacts between cells and having cells that are sufficiently stiff are two vital components of the model that ensure correct spongy mesophyll development. Without either feature, cell buckling leads to compaction rather than expansion in response to cell growth. 

The developmental trajectories produced by our model answer key questions about mesophyll development and airspace formation.  First, mesophyll porosity can develop without cell death, relying entirely on cell expansion (expansigeny)~\citep{physio:ZhangPlantCell2021}.  The ability of our model to recapitulate \emph{Arabidopsis thaliana} mesophyll development suggests that selective cell death is not required to generate developmental increases in mesophyll porosity, perhaps in contrast to airspace formation in other plant tissues~\citep{plant:WhitewoodsPLOSBIO2021}.  Second, localizing cell wall growth to regions adjacent to airspace allows for both increasing cell asphericity and the expansion of intercellular airspace.  If cell growth were constant around the perimeter of cells, then cell shape change \emph{per se} would be driven solely by mechanical stresses bending and stretching the cell wall.  That cell wall expansion is largely isolated to regions of the cell perimeter bordering airspace enables much more dramatic changes in cell shape during development~\citep{physio:ZhangPlantCell2021}. Third, we find that maintaining some adhesive bonds between cells and allowing others to break are vital for tissue stability and properly opening pore space during development. Bond breaking is important for airspace formation, since loss of cell-cell contact allows voids to open. Our model also demonstrates that losing too many adhesive bonds compromises tissue integrity and can stall or even reverse tissue expansion. Fourth, air space formation in the mesophyll need not depend on the epidermis pulling mesophyll cells apart. By maintaining constant pressure on the developing mesophyll, our model suggests that epidermal expansion can occur at the same rate as mesophyll expansion. Thus, this relatively simple model aligns with recent insights into mesophyll development in \emph{Arabidopsis thaliana} and highlights the features that are necessary for the spongy mesophyll to develop properly.

Similar to a recent computational model of epidermal development~\citep{plants:SapalaELife2018}, our model is consistent with current understanding of the molecular underpinnings of mesophyll development. Consistent with our finding that epidermal expansion need not pull mesophyll cells apart to form the intercellular airspace, \emph{Arabidopsis thaliana} mutants that lack an epidermis nonetheless produce a porous mesophyll~\citep{plants:AbeDev2003}. This finding reiterates the role of single cell behavior and growth in driving mesophyll morphogenesis.  Other features of our model, such as the preferred bending angle ($\theta_{0}$) and cell adhesion ($\beta$), have known molecular cognates. The preferred bending angle could result from molecular processes that pattern cell growth, such as microtubules that form bands in different regions of mesophyll cells and drive cell wall reinforcement~\citep{plants:PanterisProtoplasma1993,dev:PanterisNewPhyto2005,plant:WhitewoodsPLOSBIO2021}.  In general, cells are adhered together by the middle lamella, and modification of middle lamella composition can alter cell-cell adhesion~\citep{physio:DaherFrontInPlantSci2015,physio:LionettiPhytochem2015,dev:BidhendiJExpBot2015}. For example, methyl esterification, which is controlled by pectin methyl esterase, can reduce adhesion in the middle lamella~\citep{physio:LionettiPhytochem2015,physio:DaherFrontInPlantSci2015}.  Thus, while our model of mesophyll development is only a physical model, its parameters have known molecular and physiological underpinnings.

Our sensitivity tests provide additional insights into the robustness and potential diversity of mesophyll developmental trajectories.  Many of the simulations using different parameter values produced stable structures, and some simulations produced tissues that had little-to-no buckling during development, but still possessed significant deviations in $\mathcal{A}$ compared to \emph{Arabidopsis thaliana} (e.g. the results for $\Delta a = 2\times 10^{-3}$ and $\lambda = 2$ (purple hexagon) in Fig.~\ref{fig:dadl}B).  These alternative trajectories may correspond to the developmental trajectories of the spongy mesophyll for other plant species. The spongy mesophyll is very diverse, most notably in porosity, which varies from about 25\% to 75\% among species and in cell size and shape ~\citep{plants:TherouxRancourtNewPhyt2020,physio:TherouxRancourtProcRoySocB2021,plants:BorsukNewPhyto2022}.  If different parameter values in this model can recapitulate the developmental trajectories of other species, then studies of the landscape of possible phenotypes generated by our model could elucidate the biophysical basis of mesophyll diversity.  In addition to the parameters explicitly defined in our model, other assumptions in our model could be relaxed to generate a greater diversity of developmental trajectories.  For example, our implementation of the DP model assumes constant, i.e. $\phi$-independent, growth rates. Would we obtain different dynamics for the structural properties of the spongy mesophyll if, for example, areal growth depended directly on the amount of void-facing cell surfaces? In addition to varying parameter values in the model, arresting mesophyll expansion at different points along the developmental trajectory can produce an array of different mesophyll porosities and cell shapes (Fig.~\ref{fig:model}).  Heterochronic changes, such as paedomorphosis -- when mature individuals retain traits previously seen only in immature individuals of closely related species -- commonly occur during speciation.  Our model of mesophyll development recapitulates much of the cell shape diversity seen among species~\citep{plants:BorsukNewPhyto2022}, suggesting that a single developmental trajectory can itself provide numerous possible mature phenotypes. Additional rules may be necessary for simulations to correspond to mesophyll phenotypes observed in other species, e.g. disordered, planar honeycomb lattices of cells~\citep{plants:BorsukNewPhyto2022}. Therefore, better experimental characterization of the structural changes during development for a variety of species is needed to better understand the range of possible phenotypes and to determine whether the DP model can describe them.

Our simulations do miss some quantitative details of spongy mesophyll development in \emph{Arabidopsis thaliana}. For example simulated values of $\mathcal{A}$ never quite reach the minimum value of $\mathcal{A}\approx 1.05$ observed in experimental data. The inability of the model to capture this particular feature may be due to the assumption of constant growth rates, whereas real spongy mesophyll cells may have more complex cell shape feedback mechanisms that drive  this decrease in $\mathcal{A}$. Our simulations also contain more highly coordinated cells in the more porous stages of development than those in the experimental images. For example, the most porous experimental structure that we observed (i.e. the blue circle in Fig.~\ref{fig:comp}B) has an average cell-cell coordination number $z = 3.5$. However, the simulated tissues with low $\delta\mathcal{A}_{\rm late}$ typically have $z \sim 4.5 - 5$ even though the cell shapes are nearly identical to those in experiments (Figs.~\ref{fig:comp},~\ref{fig:dadl}, and~\ref{fig:kbbe}). One explanation for this discrepancy is that cells in real leaves are stabilized by out-of-plane contacts and require fewer contacts in the paradermal plane to remain mechanically stable. Extending our 2D mesophyll model into three dimensions using deformable polyhedra~\citep{jamming:WangSM2021}, would be an advancement in understanding mesophyll development. Such an approach could incorporate the presence of other tissues, such as the palisade mesophyll, the epidermis, and veins on the development of the spongy mesophyll. Nonetheless, the success of our relatively simple 2D model in recapitulating the dynamics of spongy mesophyll development in the paradermal plane suggests that our model captures all of the fundamental biological mechanisms that control mesophyll development.

\section{Methods and Materials}\label{sec:methods}

\subsection{Plant materials and growth condition}
The seeds of transgenic 
\emph{Arabidopsis thaliana} Columbia-0 (Col-0) plants expressing RbcS1B:mKO-PIP2;1~\citep{plants:ZhangNatPlants2022} were sown on plates containing 1/2 MS salts, 1\% sucrose, and 1\% agar and stratified
at 4 $^\circ$C for 48 hours before moving to a growth cabinet set to 22 $^\circ$C, 16/8 h light/dark cycle. The first true
leaves were used for imaging after their emergence from the bases of cotyledons.

\subsection{Sample preparation and microscopy}
For live cell imaging, the cotyledons were cut off in order to expose the young first true leaves after their
emergence at 6-7 days after sowing, then the rest of seedlings were mounted in 20 $\mu$L
perfluoroperhydrophenanthrene (PP11; Sigma 56919) within chambered coverglass (Nunc; Thermo
Scientific 155360) and covered by a piece of 2-3 mm thick 0.7\% phytagel containing 1\% sucrose and 1/2
MS salts. All images were obtained via point-scan confocal microscopy (Zeiss LSM 880; Airyscan).
mKusabira-Orange (mKO) excitation was performed using a 543 nm laser and fluorescence was detected
at 555-610 nm. During the imaging intervals, the samples were kept in the growth cabinet (22 $^\circ$C, 16/8 h
light/dark cycle) inside a covered glass Petri dishes with a moist paper towel to maintain humidity,
meanwhile PP11 was added occasionally to prevent drying.

\subsection{Image processing}
Confocal Z-stacks were processed in ImageJ~\citep{fiji:SchindelinNatMethods2012} and FluoRender~\citep{fluo:WanIEEE2012}. Maximum
intensity Z-projections were generated for each timepoint and used for subsequent analysis. The first
layer of spongy mesophyll was used for analysis. When required, the inner layers were manually
removed in FluoRender.

\subsection{MicroCT imaging}
For the two densest microCT scans shown in Fig.~\ref{fig:comp}B (the source of the red diamond and green square data points), seeds of \emph{Arabidopsis thaliana} Col-0 were sown on moist filter paper in a Petri dish and stratified in a cold room at 4°C for five days. They were then left to germinate at room temperature under 200 µmol PPFD $m^{-2}$ $s^{-1}$ light. Once germinated, seedlings were transferred to individual pots of 8 cm diameter and height. Seedlings were then transferred to a climate-controlled room with 21/18 $^\circ$C day/night temperature, 60\% relative humidity, and 8 h photoperiod of 500 $\mu$mol PPFD $m^{-2}$ $s^{-1}$. Plants were fertilized regularly using a complete liquid fertilizer solution. Plants were grown for about six weeks.

 One healthy and fully grown plant was brought to the TOMCAT tomographic beamline X02DA of the Swiss Light Source at the Paul Scherrer Institute (Villigen, Switzerland). Of that plant, two leaves were included in this analysis: the youngest leaf (red diamond, Fig.~\ref{fig:comp}B) large enough to be sampled (about 10-20 mm$^2$), as well as an older leaf (green square, Fig.~\ref{fig:comp}B). Before scanning, the petiole of each leaf was cut into a thin strip 1.5 mm wide and up to 1.5 cm long in between apparent higher order veins and immediately wrapped in polyimide tape~\citep{physio:TherouxRancourtProcRoySocB2021}. Each strip was scanned within 15 minutes of being prepared by imaging 1801 projections of 100 ms under a beam energy of 21 keV and magnified using a 20x objective, yielding a final voxel size of 0.325 $\mu$m (field of view: 832x832x624 µm). Projections were reconstructed using phase contrast enhancement~\citep{microct:paganinJMicro2002}, which provides a high contrast between the airspace and the mesophyll cells, sufficient to segment using a simple gray-value threshold.
 
 For the most porous microCT scan shown in Fig.~\ref{fig:comp}B (the source of the blue circle), seeds of \emph{Arabidopsis thaliana} Col-0 were plated onto wet filter paper and placed in a fridge at 4$^\circ$C for cold treatment. One week later, seeds were moved to soil (Pro-Mix BX with biofungicide; Premier Horticulture Ltd., Rivière-du-Loup, Quebec, Canada) and grown in growth chambers at Yale University (22$^\circ$C, 60\% humidity, 16h photoperiod with 100 $\mu$mols m-2 sec-1 light). Plants were watered twice per week with fertilized water (Jack’s 20-10-20 Peat-lite special; 200 ppm N). A fully expanded leaf was excised from a Col-0 plant and placed inside a plastic bag with wet paper towels, then transported to the Advanced Light Source (ALS) at the Lawrence Berkeley National Lab (Berkeley, CA). A sample was prepared by cutting out a small ($\sim$ 3 mm wide and $\sim$ 10 mm long) section at the midpoint of the leaf halfway between the midrib and leaf margin and mounting this section between pieces of Kapton Tape (E.I. du Pont de Nemours and Company, Wilmington, DE, USA). Preparation was performed less than 30 minutes prior to scanning to avoid desiccation. The sample was mounted in a sample holder and scanned at the ALS using continuous tomography mode with a 10x objective lens, capturing 1,025 projection images at 21 keV. An image stack was reconstructed following previous work~\citep{physio:TherouxRancourtNewPhyto2017}. The reconstructed stack was cropped to remove areas damaged by cutting or dehydration.

\subsection{Manual cell segmentation}
\begin{figure}[!t]
    \centering
    \includegraphics[width=0.75\textwidth]{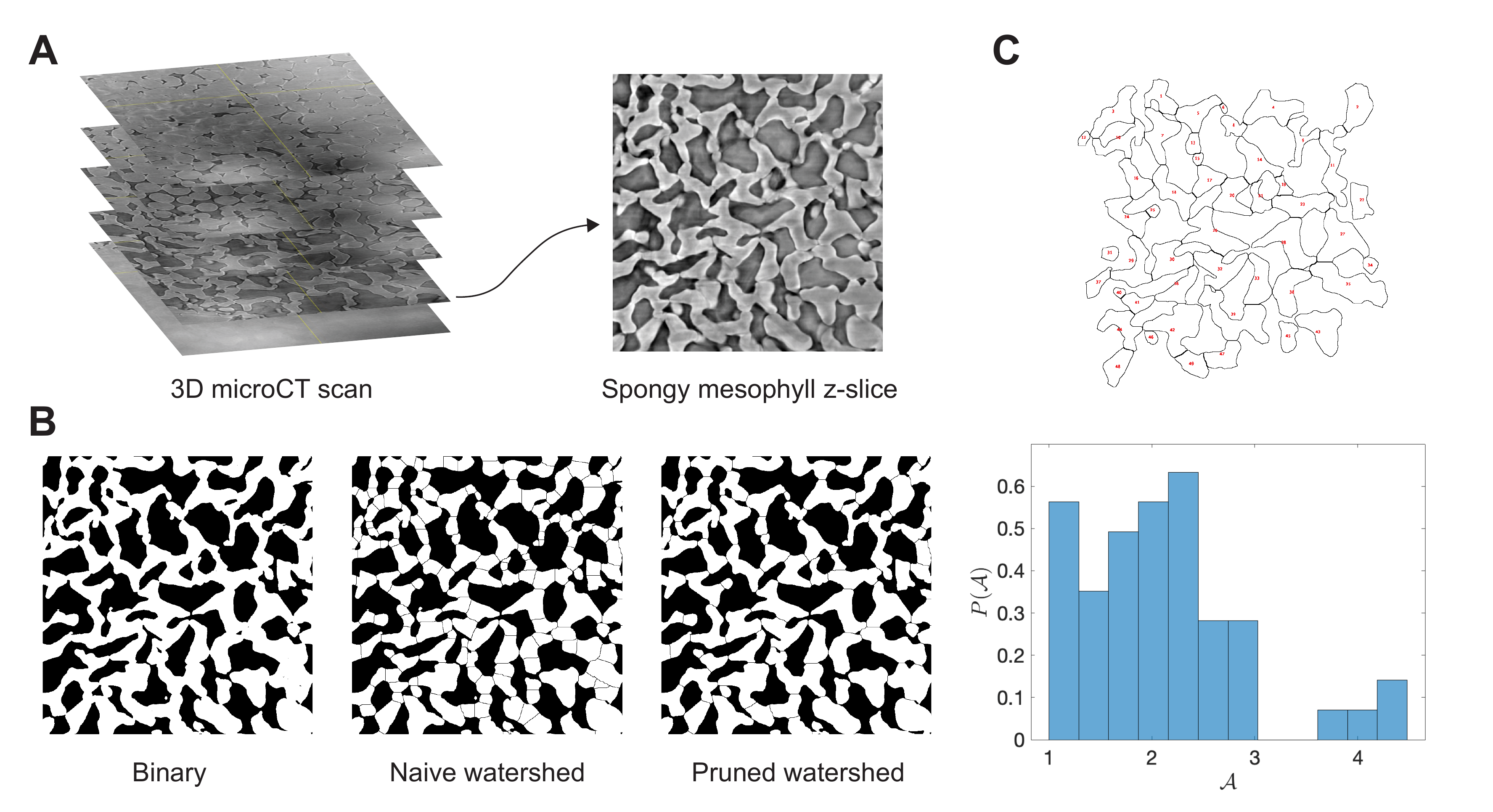}
    \caption{Process of segmenting spongy mesophyll cells from the original greyscale microCT data. (A) First, a single paradermal z-slice that represents the spongy mesophyll is selected. Here, we select the  z-slice shown in Fig.~\ref{fig:exp}D. (B) We then binarize, watershed, and prune the image to obtain approximate boundaries for the 2D cell projections. (C) We show the segmented cells, as well as the distribution of cell shape parameters $P(\mathcal{A})$. The mean of this distribution is $\overline{\mathcal{A}} = 2.13$. }
    \label{fig:segment}
\end{figure}

For each microCT data set shown in Fig.~\ref{fig:comp}, we perform a binarization and watershed procedure in ImageJ as outlined in Fig.~\ref{fig:segment}.  Once a representative $z$-slice of the spongy mesophyll is selected from a fully 3D microCT scan, we binarize the image such that all cell material takes on the same pixel value and use a watershed algorithm~\citep{watershed:VincentIEEETransPatt1991} to identify cell boundaries. However, we find that the initial boundaries drawn by the watershed algorithm divide some highly concave cells in regions where no clear boundaries can be seen in the original greyscale image. Therefore, we manually prune extraneous boundaries by comparing each naive watershed boundary with the greyscale images, and flipping void pixels to cell pixels for any incorrect boundaries. Once all extraneous boundaries are removed, we compute the shape parameter $\mathcal{A} = p^2/4\pi a$ for each remaining binary region of cell pixels with perimeters $p$ and areas $a$. In Fig.~\ref{fig:segment}, we show the distribution of cell shapes gleaned by this method from the most porous experimental snapshot (i.e. the blue circle in Fig.~\ref{fig:comp}B). 

\section{Acknowledgments}

J.D.T., A.B.R., C.R.B. and C.S.O. were supported by NSF grant no. BMMB-2029756. G.T.R. was supported by the Austrian Science Fund (FWF), projects M224, and by the Vienna Science and Technology Fund (WWTF), project LS19-013. This work was also supported by the High Performance Computing facilities operated by Yale’s Center for Research Computing. We would like to thank Dr. Anne Bonnin of the TOMCAT beamline X02DA of the Swiss Light Source, as well as Mona Nazari and Dr. Leila Fletcher for assistance during microCT sample preparation and scanning.
\bibliography{mesosim}

\begin{thebibliography}{58}%
\makeatletter
\providecommand \@ifxundefined [1]{%
 \@ifx{#1\undefined}
}%
\providecommand \@ifnum [1]{%
 \ifnum #1\expandafter \@firstoftwo
 \else \expandafter \@secondoftwo
 \fi
}%
\providecommand \@ifx [1]{%
 \ifx #1\expandafter \@firstoftwo
 \else \expandafter \@secondoftwo
 \fi
}%
\providecommand \natexlab [1]{#1}%
\providecommand \enquote  [1]{``#1''}%
\providecommand \bibnamefont  [1]{#1}%
\providecommand \bibfnamefont [1]{#1}%
\providecommand \citenamefont [1]{#1}%
\providecommand \href@noop [0]{\@secondoftwo}%
\providecommand \href [0]{\begingroup \@sanitize@url \@href}%
\providecommand \@href[1]{\@@startlink{#1}\@@href}%
\providecommand \@@href[1]{\endgroup#1\@@endlink}%
\providecommand \@sanitize@url [0]{\catcode `\\12\catcode `\$12\catcode
  `\&12\catcode `\#12\catcode `\^12\catcode `\_12\catcode `\%12\relax}%
\providecommand \@@startlink[1]{}%
\providecommand \@@endlink[0]{}%
\providecommand \url  [0]{\begingroup\@sanitize@url \@url }%
\providecommand \@url [1]{\endgroup\@href {#1}{\urlprefix }}%
\providecommand \urlprefix  [0]{URL }%
\providecommand \Eprint [0]{\href }%
\providecommand \doibase [0]{http://dx.doi.org/}%
\providecommand \selectlanguage [0]{\@gobble}%
\providecommand \bibinfo  [0]{\@secondoftwo}%
\providecommand \bibfield  [0]{\@secondoftwo}%
\providecommand \translation [1]{[#1]}%
\providecommand \BibitemOpen [0]{}%
\providecommand \bibitemStop [0]{}%
\providecommand \bibitemNoStop [0]{.\EOS\space}%
\providecommand \EOS [0]{\spacefactor3000\relax}%
\providecommand \BibitemShut  [1]{\csname bibitem#1\endcsname}%
\let\auto@bib@innerbib\@empty
\bibitem [{\citenamefont {Avery}(1933)}]{physio:AveryAmJBot1993}%
  \BibitemOpen
  \bibfield  {author} {\bibinfo {author} {\bibfnamefont {G.~S.}\ \bibnamefont
  {Avery}},\ }\href@noop {} {\bibfield  {journal} {\bibinfo  {journal}
  {American Journal of Botany}\ }\textbf {\bibinfo {volume} {20}},\ \bibinfo
  {pages} {565} (\bibinfo {year} {1933})}\BibitemShut {NoStop}%
\bibitem [{\citenamefont {Green}(1965)}]{shape:GreenJCellBio1965}%
  \BibitemOpen
  \bibfield  {author} {\bibinfo {author} {\bibfnamefont {P.~B.}\ \bibnamefont
  {Green}},\ }\href@noop {} {\bibfield  {journal} {\bibinfo  {journal} {Journal
  of Cell Biology}\ }\textbf {\bibinfo {volume} {27}},\ \bibinfo {pages} {343}
  (\bibinfo {year} {1965})}\BibitemShut {NoStop}%
\bibitem [{\citenamefont {Seago}\ \emph {et~al.}(2005)\citenamefont {Seago},
  \citenamefont {Marsh}, \citenamefont {Stevens}, \citenamefont {Soukup},
  \citenamefont {Votrubov{\'a}},\ and\ \citenamefont
  {Enstone}}]{physio:SeagoAnnBot2005}%
  \BibitemOpen
  \bibfield  {author} {\bibinfo {author} {\bibfnamefont {J.}~\bibnamefont
  {Seago}, \bibfnamefont {James~L.}}, \bibinfo {author} {\bibfnamefont {L.~C.}\
  \bibnamefont {Marsh}}, \bibinfo {author} {\bibfnamefont {K.~J.}\ \bibnamefont
  {Stevens}}, \bibinfo {author} {\bibfnamefont {A.}~\bibnamefont {Soukup}},
  \bibinfo {author} {\bibfnamefont {O.}~\bibnamefont {Votrubov{\'a}}}, \ and\
  \bibinfo {author} {\bibfnamefont {D.~E.}\ \bibnamefont {Enstone}},\
  }\href@noop {} {\bibfield  {journal} {\bibinfo  {journal} {Annals of Botany}\
  }\textbf {\bibinfo {volume} {96}},\ \bibinfo {pages} {565} (\bibinfo {year}
  {2005})}\BibitemShut {NoStop}%
\bibitem [{\citenamefont {Ivakov}\ and\ \citenamefont
  {Persson}(2013)}]{shape:IvakovFrontInPlantSci2013}%
  \BibitemOpen
  \bibfield  {author} {\bibinfo {author} {\bibfnamefont {A.}~\bibnamefont
  {Ivakov}}\ and\ \bibinfo {author} {\bibfnamefont {S.}~\bibnamefont
  {Persson}},\ }\href@noop {} {\bibfield  {journal} {\bibinfo  {journal}
  {Frontiers in Plant Science}\ }\textbf {\bibinfo {volume} {4}} (\bibinfo
  {year} {2013})}\BibitemShut {NoStop}%
\bibitem [{\citenamefont {Sapala}\ \emph {et~al.}(2018)\citenamefont {Sapala},
  \citenamefont {Runions}, \citenamefont {Routier-Kierzkowska}, \citenamefont
  {Das~Gupta}, \citenamefont {Hong}, \citenamefont {Hofhuis}, \citenamefont
  {Verger}, \citenamefont {Mosca}, \citenamefont {Li}, \citenamefont {Hay},
  \citenamefont {Hamant}, \citenamefont {Roeder}, \citenamefont {Tsiantis},
  \citenamefont {Prusinkiewicz},\ and\ \citenamefont
  {Smith}}]{plants:SapalaELife2018}%
  \BibitemOpen
  \bibfield  {author} {\bibinfo {author} {\bibfnamefont {A.}~\bibnamefont
  {Sapala}}, \bibinfo {author} {\bibfnamefont {A.}~\bibnamefont {Runions}},
  \bibinfo {author} {\bibfnamefont {A.-L.}\ \bibnamefont
  {Routier-Kierzkowska}}, \bibinfo {author} {\bibfnamefont {M.}~\bibnamefont
  {Das~Gupta}}, \bibinfo {author} {\bibfnamefont {L.}~\bibnamefont {Hong}},
  \bibinfo {author} {\bibfnamefont {H.}~\bibnamefont {Hofhuis}}, \bibinfo
  {author} {\bibfnamefont {S.}~\bibnamefont {Verger}}, \bibinfo {author}
  {\bibfnamefont {G.}~\bibnamefont {Mosca}}, \bibinfo {author} {\bibfnamefont
  {C.-B.}\ \bibnamefont {Li}}, \bibinfo {author} {\bibfnamefont
  {A.}~\bibnamefont {Hay}}, \bibinfo {author} {\bibfnamefont {O.}~\bibnamefont
  {Hamant}}, \bibinfo {author} {\bibfnamefont {A.~H.}\ \bibnamefont {Roeder}},
  \bibinfo {author} {\bibfnamefont {M.}~\bibnamefont {Tsiantis}}, \bibinfo
  {author} {\bibfnamefont {P.}~\bibnamefont {Prusinkiewicz}}, \ and\ \bibinfo
  {author} {\bibfnamefont {R.~S.}\ \bibnamefont {Smith}},\ }\href@noop {}
  {\bibfield  {journal} {\bibinfo  {journal} {eLife}\ }\textbf {\bibinfo
  {volume} {7}},\ \bibinfo {pages} {e32794} (\bibinfo {year}
  {2018})}\BibitemShut {NoStop}%
\bibitem [{\citenamefont {Antreich}\ \emph {et~al.}(2019)\citenamefont
  {Antreich}, \citenamefont {Xiao}, \citenamefont {Huss}, \citenamefont
  {Horbelt}, \citenamefont {Eder}, \citenamefont {Weinkamer},\ and\
  \citenamefont {Gierlinger}}]{plants:AntreichAdvSci2019}%
  \BibitemOpen
  \bibfield  {author} {\bibinfo {author} {\bibfnamefont {S.~J.}\ \bibnamefont
  {Antreich}}, \bibinfo {author} {\bibfnamefont {N.}~\bibnamefont {Xiao}},
  \bibinfo {author} {\bibfnamefont {J.~C.}\ \bibnamefont {Huss}}, \bibinfo
  {author} {\bibfnamefont {N.}~\bibnamefont {Horbelt}}, \bibinfo {author}
  {\bibfnamefont {M.}~\bibnamefont {Eder}}, \bibinfo {author} {\bibfnamefont
  {R.}~\bibnamefont {Weinkamer}}, \ and\ \bibinfo {author} {\bibfnamefont
  {N.}~\bibnamefont {Gierlinger}},\ }\href@noop {} {\bibfield  {journal}
  {\bibinfo  {journal} {Advanced Science}\ }\textbf {\bibinfo {volume} {6}},\
  \bibinfo {pages} {1900644} (\bibinfo {year} {2019})}\BibitemShut {NoStop}%
\bibitem [{\citenamefont {Roddy}\ \emph {et~al.}(2020)\citenamefont {Roddy},
  \citenamefont {Th{\'e}roux-Rancourt}, \citenamefont {Abbo}, \citenamefont
  {Benedetti}, \citenamefont {Brodersen}, \citenamefont {Castro}, \citenamefont
  {Castro}, \citenamefont {Gilbride}, \citenamefont {Jensen}, \citenamefont
  {Jiang}, \citenamefont {Perkins}, \citenamefont {Perkins}, \citenamefont
  {Loureiro}, \citenamefont {Syed}, \citenamefont {Thompson}, \citenamefont
  {Kuebbing},\ and\ \citenamefont {Simonin}}]{plantevo:RoddyIJPS2020}%
  \BibitemOpen
  \bibfield  {author} {\bibinfo {author} {\bibfnamefont {A.~B.}\ \bibnamefont
  {Roddy}}, \bibinfo {author} {\bibfnamefont {G.}~\bibnamefont
  {Th{\'e}roux-Rancourt}}, \bibinfo {author} {\bibfnamefont {T.}~\bibnamefont
  {Abbo}}, \bibinfo {author} {\bibfnamefont {J.~W.}\ \bibnamefont {Benedetti}},
  \bibinfo {author} {\bibfnamefont {C.~R.}\ \bibnamefont {Brodersen}}, \bibinfo
  {author} {\bibfnamefont {M.}~\bibnamefont {Castro}}, \bibinfo {author}
  {\bibfnamefont {S.}~\bibnamefont {Castro}}, \bibinfo {author} {\bibfnamefont
  {A.~B.}\ \bibnamefont {Gilbride}}, \bibinfo {author} {\bibfnamefont
  {B.}~\bibnamefont {Jensen}}, \bibinfo {author} {\bibfnamefont {G.-F.}\
  \bibnamefont {Jiang}}, \bibinfo {author} {\bibfnamefont {J.~A.}\ \bibnamefont
  {Perkins}}, \bibinfo {author} {\bibfnamefont {S.~D.}\ \bibnamefont
  {Perkins}}, \bibinfo {author} {\bibfnamefont {J.}~\bibnamefont {Loureiro}},
  \bibinfo {author} {\bibfnamefont {Z.}~\bibnamefont {Syed}}, \bibinfo {author}
  {\bibfnamefont {R.~A.}\ \bibnamefont {Thompson}}, \bibinfo {author}
  {\bibfnamefont {S.~E.}\ \bibnamefont {Kuebbing}}, \ and\ \bibinfo {author}
  {\bibfnamefont {K.~A.}\ \bibnamefont {Simonin}},\ }\href@noop {} {\bibfield
  {journal} {\bibinfo  {journal} {International Journal of Plant Sciences}\
  }\textbf {\bibinfo {volume} {181}},\ \bibinfo {pages} {75} (\bibinfo {year}
  {2020})}\BibitemShut {NoStop}%
\bibitem [{\citenamefont {Hamant}\ \emph {et~al.}(2008)\citenamefont {Hamant},
  \citenamefont {Heisler}, \citenamefont {J{\"o}nsson}, \citenamefont
  {Krupinski}, \citenamefont {Uyttewaal}, \citenamefont {Bokov}, \citenamefont
  {Corson}, \citenamefont {Sahlin}, \citenamefont {Boudaoud}, \citenamefont
  {Meyerowitz}, \citenamefont {Couder},\ and\ \citenamefont
  {Traas}}]{physio:HamantScience2008}%
  \BibitemOpen
  \bibfield  {author} {\bibinfo {author} {\bibfnamefont {O.}~\bibnamefont
  {Hamant}}, \bibinfo {author} {\bibfnamefont {M.~G.}\ \bibnamefont {Heisler}},
  \bibinfo {author} {\bibfnamefont {H.}~\bibnamefont {J{\"o}nsson}}, \bibinfo
  {author} {\bibfnamefont {P.}~\bibnamefont {Krupinski}}, \bibinfo {author}
  {\bibfnamefont {M.}~\bibnamefont {Uyttewaal}}, \bibinfo {author}
  {\bibfnamefont {P.}~\bibnamefont {Bokov}}, \bibinfo {author} {\bibfnamefont
  {F.}~\bibnamefont {Corson}}, \bibinfo {author} {\bibfnamefont
  {P.}~\bibnamefont {Sahlin}}, \bibinfo {author} {\bibfnamefont
  {A.}~\bibnamefont {Boudaoud}}, \bibinfo {author} {\bibfnamefont {E.~M.}\
  \bibnamefont {Meyerowitz}}, \bibinfo {author} {\bibfnamefont
  {Y.}~\bibnamefont {Couder}}, \ and\ \bibinfo {author} {\bibfnamefont
  {J.}~\bibnamefont {Traas}},\ }\href@noop {} {\bibfield  {journal} {\bibinfo
  {journal} {Science}\ }\textbf {\bibinfo {volume} {322}},\ \bibinfo {pages}
  {1650} (\bibinfo {year} {2008})}\BibitemShut {NoStop}%
\bibitem [{\citenamefont {Sampathkumar}\ \emph {et~al.}(2014)\citenamefont
  {Sampathkumar}, \citenamefont {Yan}, \citenamefont {Krupinski},\ and\
  \citenamefont {Meyerowitz}}]{physio:SampathkumarCurrBio2014}%
  \BibitemOpen
  \bibfield  {author} {\bibinfo {author} {\bibfnamefont {A.}~\bibnamefont
  {Sampathkumar}}, \bibinfo {author} {\bibfnamefont {A.}~\bibnamefont {Yan}},
  \bibinfo {author} {\bibfnamefont {P.}~\bibnamefont {Krupinski}}, \ and\
  \bibinfo {author} {\bibfnamefont {E.~M.}\ \bibnamefont {Meyerowitz}},\
  }\href@noop {} {\bibfield  {journal} {\bibinfo  {journal} {Current Biology}\
  }\textbf {\bibinfo {volume} {24}},\ \bibinfo {pages} {R475} (\bibinfo {year}
  {2014})}\BibitemShut {NoStop}%
\bibitem [{\citenamefont {V\H{o}f\'{e}ly}\ \emph {et~al.}(2019)\citenamefont
  {V\H{o}f\'{e}ly}, \citenamefont {Gallagher}, \citenamefont {Pisano},
  \citenamefont {Bartlett},\ and\ \citenamefont
  {Braybrook}}]{plants:VofelyNewPhyto2019}%
  \BibitemOpen
  \bibfield  {author} {\bibinfo {author} {\bibfnamefont {R.~V.}\ \bibnamefont
  {V\H{o}f\'{e}ly}}, \bibinfo {author} {\bibfnamefont {J.}~\bibnamefont
  {Gallagher}}, \bibinfo {author} {\bibfnamefont {G.~D.}\ \bibnamefont
  {Pisano}}, \bibinfo {author} {\bibfnamefont {M.}~\bibnamefont {Bartlett}}, \
  and\ \bibinfo {author} {\bibfnamefont {S.~A.}\ \bibnamefont {Braybrook}},\
  }\href@noop {} {\bibfield  {journal} {\bibinfo  {journal} {New Phytologist}\
  }\textbf {\bibinfo {volume} {221}},\ \bibinfo {pages} {540} (\bibinfo {year}
  {2019})}\BibitemShut {NoStop}%
\bibitem [{\citenamefont {Roeder}\ \emph {et~al.}(2011)\citenamefont {Roeder},
  \citenamefont {Tarr}, \citenamefont {Tobin}, \citenamefont {Zhang},
  \citenamefont {Chickarmane}, \citenamefont {Cunha},\ and\ \citenamefont
  {Meyerowitz}}]{plants:RoederNatRevMCB2011}%
  \BibitemOpen
  \bibfield  {author} {\bibinfo {author} {\bibfnamefont {A.~H.~K.}\
  \bibnamefont {Roeder}}, \bibinfo {author} {\bibfnamefont {P.~T.}\
  \bibnamefont {Tarr}}, \bibinfo {author} {\bibfnamefont {C.}~\bibnamefont
  {Tobin}}, \bibinfo {author} {\bibfnamefont {X.}~\bibnamefont {Zhang}},
  \bibinfo {author} {\bibfnamefont {V.}~\bibnamefont {Chickarmane}}, \bibinfo
  {author} {\bibfnamefont {A.}~\bibnamefont {Cunha}}, \ and\ \bibinfo {author}
  {\bibfnamefont {E.~M.}\ \bibnamefont {Meyerowitz}},\ }\href@noop {}
  {\bibfield  {journal} {\bibinfo  {journal} {Nature Reviews Molecular Cell
  Biology}\ }\textbf {\bibinfo {volume} {12}},\ \bibinfo {pages} {265}
  (\bibinfo {year} {2011})}\BibitemShut {NoStop}%
\bibitem [{\citenamefont {Boudon}\ \emph {et~al.}(2015)\citenamefont {Boudon},
  \citenamefont {Chopard}, \citenamefont {Ali}, \citenamefont {Gilles},
  \citenamefont {Hamant}, \citenamefont {Boudaoud}, \citenamefont {Traas},\
  and\ \citenamefont {Godin}}]{cellmodel:BoudonPLOSCompBio2015}%
  \BibitemOpen
  \bibfield  {author} {\bibinfo {author} {\bibfnamefont {F.}~\bibnamefont
  {Boudon}}, \bibinfo {author} {\bibfnamefont {J.}~\bibnamefont {Chopard}},
  \bibinfo {author} {\bibfnamefont {O.}~\bibnamefont {Ali}}, \bibinfo {author}
  {\bibfnamefont {B.}~\bibnamefont {Gilles}}, \bibinfo {author} {\bibfnamefont
  {O.}~\bibnamefont {Hamant}}, \bibinfo {author} {\bibfnamefont
  {A.}~\bibnamefont {Boudaoud}}, \bibinfo {author} {\bibfnamefont
  {J.}~\bibnamefont {Traas}}, \ and\ \bibinfo {author} {\bibfnamefont
  {C.}~\bibnamefont {Godin}},\ }\href@noop {} {\bibfield  {journal} {\bibinfo
  {journal} {PLOS Computational Biology}\ }\textbf {\bibinfo {volume} {11}},\
  \bibinfo {pages} {1} (\bibinfo {year} {2015})}\BibitemShut {NoStop}%
\bibitem [{\citenamefont {Lundgren}\ \emph {et~al.}(2019)\citenamefont
  {Lundgren}, \citenamefont {Mathers}, \citenamefont {Baillie}, \citenamefont
  {Dunn}, \citenamefont {Wilson}, \citenamefont {Hunt}, \citenamefont {Pajor},
  \citenamefont {Fradera-Soler}, \citenamefont {Rolfe}, \citenamefont
  {Osborne}, \citenamefont {Sturrock}, \citenamefont {Gray}, \citenamefont
  {Mooney},\ and\ \citenamefont {Fleming}}]{plants:LundgrenNatComm2019}%
  \BibitemOpen
  \bibfield  {author} {\bibinfo {author} {\bibfnamefont {M.~R.}\ \bibnamefont
  {Lundgren}}, \bibinfo {author} {\bibfnamefont {A.}~\bibnamefont {Mathers}},
  \bibinfo {author} {\bibfnamefont {A.~L.}\ \bibnamefont {Baillie}}, \bibinfo
  {author} {\bibfnamefont {J.}~\bibnamefont {Dunn}}, \bibinfo {author}
  {\bibfnamefont {M.~J.}\ \bibnamefont {Wilson}}, \bibinfo {author}
  {\bibfnamefont {L.}~\bibnamefont {Hunt}}, \bibinfo {author} {\bibfnamefont
  {R.}~\bibnamefont {Pajor}}, \bibinfo {author} {\bibfnamefont
  {M.}~\bibnamefont {Fradera-Soler}}, \bibinfo {author} {\bibfnamefont
  {S.}~\bibnamefont {Rolfe}}, \bibinfo {author} {\bibfnamefont {C.~P.}\
  \bibnamefont {Osborne}}, \bibinfo {author} {\bibfnamefont {C.~J.}\
  \bibnamefont {Sturrock}}, \bibinfo {author} {\bibfnamefont {J.~E.}\
  \bibnamefont {Gray}}, \bibinfo {author} {\bibfnamefont {S.~J.}\ \bibnamefont
  {Mooney}}, \ and\ \bibinfo {author} {\bibfnamefont {A.~J.}\ \bibnamefont
  {Fleming}},\ }\href@noop {} {\bibfield  {journal} {\bibinfo  {journal}
  {Nature Communications}\ }\textbf {\bibinfo {volume} {10}},\ \bibinfo {pages}
  {2825} (\bibinfo {year} {2019})}\BibitemShut {NoStop}%
\bibitem [{\citenamefont {Th{\'e}roux-Rancourt}\ \emph
  {et~al.}(2021)\citenamefont {Th{\'e}roux-Rancourt}, \citenamefont {Roddy},
  \citenamefont {Earles}, \citenamefont {Gilbert}, \citenamefont {Zwieniecki},
  \citenamefont {Boyce}, \citenamefont {Tholen}, \citenamefont {McElrone},
  \citenamefont {Simonin},\ and\ \citenamefont
  {Brodersen}}]{physio:TherouxRancourtProcRoySocB2021}%
  \BibitemOpen
  \bibfield  {author} {\bibinfo {author} {\bibfnamefont {G.}~\bibnamefont
  {Th{\'e}roux-Rancourt}}, \bibinfo {author} {\bibfnamefont {A.~B.}\
  \bibnamefont {Roddy}}, \bibinfo {author} {\bibfnamefont {J.~M.}\ \bibnamefont
  {Earles}}, \bibinfo {author} {\bibfnamefont {M.~E.}\ \bibnamefont {Gilbert}},
  \bibinfo {author} {\bibfnamefont {M.~A.}\ \bibnamefont {Zwieniecki}},
  \bibinfo {author} {\bibfnamefont {C.~K.}\ \bibnamefont {Boyce}}, \bibinfo
  {author} {\bibfnamefont {D.}~\bibnamefont {Tholen}}, \bibinfo {author}
  {\bibfnamefont {A.~J.}\ \bibnamefont {McElrone}}, \bibinfo {author}
  {\bibfnamefont {K.~A.}\ \bibnamefont {Simonin}}, \ and\ \bibinfo {author}
  {\bibfnamefont {C.~R.}\ \bibnamefont {Brodersen}},\ }\href@noop {} {\bibfield
   {journal} {\bibinfo  {journal} {Proceedings of the Royal Society B:
  Biological Sciences}\ }\textbf {\bibinfo {volume} {288}},\ \bibinfo {pages}
  {20203145} (\bibinfo {year} {2021})}\BibitemShut {NoStop}%
\bibitem [{\citenamefont {Scott}\ \emph {et~al.}(1948)\citenamefont {Scott},
  \citenamefont {Schroeder},\ and\ \citenamefont
  {Turrell}}]{plants:ScottBotGaz1948}%
  \BibitemOpen
  \bibfield  {author} {\bibinfo {author} {\bibfnamefont {F.~M.}\ \bibnamefont
  {Scott}}, \bibinfo {author} {\bibfnamefont {M.~R.}\ \bibnamefont
  {Schroeder}}, \ and\ \bibinfo {author} {\bibfnamefont {F.~M.}\ \bibnamefont
  {Turrell}},\ }\href@noop {} {\bibfield  {journal} {\bibinfo  {journal}
  {Botanical Gazette}\ }\textbf {\bibinfo {volume} {109}},\ \bibinfo {pages}
  {381} (\bibinfo {year} {1948})}\BibitemShut {NoStop}%
\bibitem [{\citenamefont {Borsuk}\ \emph {et~al.}(2022)\citenamefont {Borsuk},
  \citenamefont {Roddy}, \citenamefont {Th{\'e}roux-Rancourt},\ and\
  \citenamefont {Brodersen}}]{plants:BorsukNewPhyto2022}%
  \BibitemOpen
  \bibfield  {author} {\bibinfo {author} {\bibfnamefont {A.~M.}\ \bibnamefont
  {Borsuk}}, \bibinfo {author} {\bibfnamefont {A.~B.}\ \bibnamefont {Roddy}},
  \bibinfo {author} {\bibfnamefont {G.}~\bibnamefont {Th{\'e}roux-Rancourt}}, \
  and\ \bibinfo {author} {\bibfnamefont {C.~R.}\ \bibnamefont {Brodersen}},\
  }\href@noop {} {\bibfield  {journal} {\bibinfo  {journal} {New Phytologist}\
  } (\bibinfo {year} {2022})}\BibitemShut {NoStop}%
\bibitem [{\citenamefont {{Th{\'e}roux-Rancourt}}\ \emph
  {et~al.}(2020)\citenamefont {{Th{\'e}roux-Rancourt}}, \citenamefont
  {Voggeneder},\ and\ \citenamefont
  {Tholen}}]{plants:TherouxRancourtNewPhyt2020}%
  \BibitemOpen
  \bibfield  {author} {\bibinfo {author} {\bibfnamefont {G.}~\bibnamefont
  {{Th{\'e}roux-Rancourt}}}, \bibinfo {author} {\bibfnamefont {K.}~\bibnamefont
  {Voggeneder}}, \ and\ \bibinfo {author} {\bibfnamefont {D.}~\bibnamefont
  {Tholen}},\ }\href@noop {} {\bibfield  {journal} {\bibinfo  {journal} {New
  Phytologist}\ }\textbf {\bibinfo {volume} {225}},\ \bibinfo {pages} {2239}
  (\bibinfo {year} {2020})}\BibitemShut {NoStop}%
\bibitem [{\citenamefont {Terashima}\ \emph {et~al.}(2001)\citenamefont
  {Terashima}, \citenamefont {Miyazawa},\ and\ \citenamefont
  {Hanba}}]{plants:TerashimaJPlantResearch2001}%
  \BibitemOpen
  \bibfield  {author} {\bibinfo {author} {\bibfnamefont {I.}~\bibnamefont
  {Terashima}}, \bibinfo {author} {\bibfnamefont {S.-I.}\ \bibnamefont
  {Miyazawa}}, \ and\ \bibinfo {author} {\bibfnamefont {Y.~T.}\ \bibnamefont
  {Hanba}},\ }\href@noop {} {\bibfield  {journal} {\bibinfo  {journal} {Journal
  of Plant Research}\ }\textbf {\bibinfo {volume} {114}},\ \bibinfo {pages}
  {93} (\bibinfo {year} {2001})}\BibitemShut {NoStop}%
\bibitem [{\citenamefont {EVANS}\ and\ \citenamefont
  {VOGELMANN}(2003)}]{plants:EvansPCE2003}%
  \BibitemOpen
  \bibfield  {author} {\bibinfo {author} {\bibfnamefont {J.~R.}\ \bibnamefont
  {EVANS}}\ and\ \bibinfo {author} {\bibfnamefont {T.~C.}\ \bibnamefont
  {VOGELMANN}},\ }\href@noop {} {\bibfield  {journal} {\bibinfo  {journal}
  {Plant, Cell \& Environment}\ }\textbf {\bibinfo {volume} {26}},\ \bibinfo
  {pages} {547} (\bibinfo {year} {2003})}\BibitemShut {NoStop}%
\bibitem [{\citenamefont {Lehmeier}\ \emph {et~al.}(2017)\citenamefont
  {Lehmeier}, \citenamefont {Pajor}, \citenamefont {Lundgren}, \citenamefont
  {Mathers}, \citenamefont {Sloan}, \citenamefont {Bauch}, \citenamefont
  {Mitchell}, \citenamefont {Bellasio}, \citenamefont {Green}, \citenamefont
  {Bouyer}, \citenamefont {Schnittger}, \citenamefont {Sturrock}, \citenamefont
  {Osborne}, \citenamefont {Rolfe}, \citenamefont {Mooney},\ and\ \citenamefont
  {Fleming}}]{plants:LehmeierThePlantJ2017}%
  \BibitemOpen
  \bibfield  {author} {\bibinfo {author} {\bibfnamefont {C.}~\bibnamefont
  {Lehmeier}}, \bibinfo {author} {\bibfnamefont {R.}~\bibnamefont {Pajor}},
  \bibinfo {author} {\bibfnamefont {M.~R.}\ \bibnamefont {Lundgren}}, \bibinfo
  {author} {\bibfnamefont {A.}~\bibnamefont {Mathers}}, \bibinfo {author}
  {\bibfnamefont {J.}~\bibnamefont {Sloan}}, \bibinfo {author} {\bibfnamefont
  {M.}~\bibnamefont {Bauch}}, \bibinfo {author} {\bibfnamefont
  {A.}~\bibnamefont {Mitchell}}, \bibinfo {author} {\bibfnamefont
  {C.}~\bibnamefont {Bellasio}}, \bibinfo {author} {\bibfnamefont
  {A.}~\bibnamefont {Green}}, \bibinfo {author} {\bibfnamefont
  {D.}~\bibnamefont {Bouyer}}, \bibinfo {author} {\bibfnamefont
  {A.}~\bibnamefont {Schnittger}}, \bibinfo {author} {\bibfnamefont
  {C.}~\bibnamefont {Sturrock}}, \bibinfo {author} {\bibfnamefont {C.~P.}\
  \bibnamefont {Osborne}}, \bibinfo {author} {\bibfnamefont {S.}~\bibnamefont
  {Rolfe}}, \bibinfo {author} {\bibfnamefont {S.}~\bibnamefont {Mooney}}, \
  and\ \bibinfo {author} {\bibfnamefont {A.~J.}\ \bibnamefont {Fleming}},\
  }\href@noop {} {\bibfield  {journal} {\bibinfo  {journal} {The Plant
  Journal}\ }\textbf {\bibinfo {volume} {92}},\ \bibinfo {pages} {981}
  (\bibinfo {year} {2017})}\BibitemShut {NoStop}%
\bibitem [{\citenamefont {Psaras}\ and\ \citenamefont
  {Rhizopoulou}(1995)}]{physio:PsarasNewPhyto1995}%
  \BibitemOpen
  \bibfield  {author} {\bibinfo {author} {\bibfnamefont {G.~K.}\ \bibnamefont
  {Psaras}}\ and\ \bibinfo {author} {\bibfnamefont {S.}~\bibnamefont
  {Rhizopoulou}},\ }\href@noop {} {\bibfield  {journal} {\bibinfo  {journal}
  {New Phytologist}\ }\textbf {\bibinfo {volume} {131}},\ \bibinfo {pages}
  {303} (\bibinfo {year} {1995})}\BibitemShut {NoStop}%
\bibitem [{\citenamefont {Panteris}\ and\ \citenamefont
  {Galatis}(2005)}]{dev:PanterisNewPhyto2005}%
  \BibitemOpen
  \bibfield  {author} {\bibinfo {author} {\bibfnamefont {E.}~\bibnamefont
  {Panteris}}\ and\ \bibinfo {author} {\bibfnamefont {B.}~\bibnamefont
  {Galatis}},\ }\href@noop {} {\bibfield  {journal} {\bibinfo  {journal} {New
  Phytologist}\ }\textbf {\bibinfo {volume} {167}},\ \bibinfo {pages} {721}
  (\bibinfo {year} {2005})}\BibitemShut {NoStop}%
\bibitem [{\citenamefont {Zhang}\ \emph {et~al.}(2021)\citenamefont {Zhang},
  \citenamefont {McEvoy}, \citenamefont {Le},\ and\ \citenamefont
  {Ambrose}}]{physio:ZhangPlantCell2021}%
  \BibitemOpen
  \bibfield  {author} {\bibinfo {author} {\bibfnamefont {L.}~\bibnamefont
  {Zhang}}, \bibinfo {author} {\bibfnamefont {D.}~\bibnamefont {McEvoy}},
  \bibinfo {author} {\bibfnamefont {Y.}~\bibnamefont {Le}}, \ and\ \bibinfo
  {author} {\bibfnamefont {C.}~\bibnamefont {Ambrose}},\ }\href@noop {}
  {\bibfield  {journal} {\bibinfo  {journal} {The Plant Cell}\ }\textbf
  {\bibinfo {volume} {33}},\ \bibinfo {pages} {623} (\bibinfo {year}
  {2021})}\BibitemShut {NoStop}%
\bibitem [{\citenamefont {Whitewoods}(2021)}]{plant:WhitewoodsPLOSBIO2021}%
  \BibitemOpen
  \bibfield  {author} {\bibinfo {author} {\bibfnamefont {C.~D.}\ \bibnamefont
  {Whitewoods}},\ }\href@noop {} {\bibfield  {journal} {\bibinfo  {journal}
  {PLOS Biology}\ }\textbf {\bibinfo {volume} {19}},\ \bibinfo {pages} {1}
  (\bibinfo {year} {2021})}\BibitemShut {NoStop}%
\bibitem [{\citenamefont {Wuyts}\ \emph {et~al.}(2010)\citenamefont {Wuyts},
  \citenamefont {Palauqui}, \citenamefont {Conejero}, \citenamefont {Verdeil},
  \citenamefont {Granier},\ and\ \citenamefont
  {Massonnet}}]{physio:WuytsPlantMethods2010}%
  \BibitemOpen
  \bibfield  {author} {\bibinfo {author} {\bibfnamefont {N.}~\bibnamefont
  {Wuyts}}, \bibinfo {author} {\bibfnamefont {J.-C.}\ \bibnamefont {Palauqui}},
  \bibinfo {author} {\bibfnamefont {G.}~\bibnamefont {Conejero}}, \bibinfo
  {author} {\bibfnamefont {J.-L.}\ \bibnamefont {Verdeil}}, \bibinfo {author}
  {\bibfnamefont {C.}~\bibnamefont {Granier}}, \ and\ \bibinfo {author}
  {\bibfnamefont {C.}~\bibnamefont {Massonnet}},\ }\href@noop {} {\bibfield
  {journal} {\bibinfo  {journal} {Plant Methods}\ }\textbf {\bibinfo {volume}
  {6}},\ \bibinfo {pages} {17} (\bibinfo {year} {2010})}\BibitemShut {NoStop}%
\bibitem [{\citenamefont {Kalve}\ \emph {et~al.}(2014)\citenamefont {Kalve},
  \citenamefont {Fotschki}, \citenamefont {Beeckman}, \citenamefont
  {Vissenberg},\ and\ \citenamefont {Beemster}}]{physio:KalveJExpBot2014}%
  \BibitemOpen
  \bibfield  {author} {\bibinfo {author} {\bibfnamefont {S.}~\bibnamefont
  {Kalve}}, \bibinfo {author} {\bibfnamefont {J.}~\bibnamefont {Fotschki}},
  \bibinfo {author} {\bibfnamefont {T.}~\bibnamefont {Beeckman}}, \bibinfo
  {author} {\bibfnamefont {K.}~\bibnamefont {Vissenberg}}, \ and\ \bibinfo
  {author} {\bibfnamefont {G.~T.~S.}\ \bibnamefont {Beemster}},\ }\href@noop {}
  {\bibfield  {journal} {\bibinfo  {journal} {Journal of Experimental Botany}\
  }\textbf {\bibinfo {volume} {65}},\ \bibinfo {pages} {6385} (\bibinfo {year}
  {2014})}\BibitemShut {NoStop}%
\bibitem [{\citenamefont {Harwood}\ \emph {et~al.}(2021)\citenamefont
  {Harwood}, \citenamefont {Th{\'e}roux-Rancourt},\ and\ \citenamefont
  {Barbour}}]{plants:HarwoodPCE2021}%
  \BibitemOpen
  \bibfield  {author} {\bibinfo {author} {\bibfnamefont {R.}~\bibnamefont
  {Harwood}}, \bibinfo {author} {\bibfnamefont {G.}~\bibnamefont
  {Th{\'e}roux-Rancourt}}, \ and\ \bibinfo {author} {\bibfnamefont {M.~M.}\
  \bibnamefont {Barbour}},\ }\href@noop {} {\bibfield  {journal} {\bibinfo
  {journal} {Plant, Cell \& Environment}\ }\textbf {\bibinfo {volume} {44}},\
  \bibinfo {pages} {2455} (\bibinfo {year} {2021})}\BibitemShut {NoStop}%
\bibitem [{\citenamefont {Daher}\ and\ \citenamefont
  {Braybrook}(2015)}]{physio:DaherFrontInPlantSci2015}%
  \BibitemOpen
  \bibfield  {author} {\bibinfo {author} {\bibfnamefont {F.~B.}\ \bibnamefont
  {Daher}}\ and\ \bibinfo {author} {\bibfnamefont {S.~A.}\ \bibnamefont
  {Braybrook}},\ }\href@noop {} {\bibfield  {journal} {\bibinfo  {journal}
  {Frontiers in Plant Science}\ }\textbf {\bibinfo {volume} {6}},\ \bibinfo
  {pages} {523} (\bibinfo {year} {2015})}\BibitemShut {NoStop}%
\bibitem [{\citenamefont {Zimmermann}\ \emph {et~al.}(1980)\citenamefont
  {Zimmermann}, \citenamefont {H{\"u}sken},\ and\ \citenamefont
  {Schulze}}]{plant:ZimmermannPlanta1980}%
  \BibitemOpen
  \bibfield  {author} {\bibinfo {author} {\bibfnamefont {U.}~\bibnamefont
  {Zimmermann}}, \bibinfo {author} {\bibfnamefont {D.}~\bibnamefont
  {H{\"u}sken}}, \ and\ \bibinfo {author} {\bibfnamefont {E.~D.}\ \bibnamefont
  {Schulze}},\ }\href@noop {} {\bibfield  {journal} {\bibinfo  {journal}
  {Planta}\ }\textbf {\bibinfo {volume} {149}},\ \bibinfo {pages} {445}
  (\bibinfo {year} {1980})}\BibitemShut {NoStop}%
\bibitem [{\citenamefont {Eng}\ and\ \citenamefont
  {Sampathkumar}(2018)}]{physio:EngCurrOpPlantBio2018}%
  \BibitemOpen
  \bibfield  {author} {\bibinfo {author} {\bibfnamefont {R.~C.}\ \bibnamefont
  {Eng}}\ and\ \bibinfo {author} {\bibfnamefont {A.}~\bibnamefont
  {Sampathkumar}},\ }\href@noop {} {\bibfield  {journal} {\bibinfo  {journal}
  {Current Opinion in Plant Biology}\ }\textbf {\bibinfo {volume} {46}},\
  \bibinfo {pages} {25 } (\bibinfo {year} {2018})}\BibitemShut {NoStop}%
\bibitem [{\citenamefont {Boromand}\ \emph {et~al.}(2018)\citenamefont
  {Boromand}, \citenamefont {Signoriello}, \citenamefont {Ye}, \citenamefont
  {O'Hern},\ and\ \citenamefont {Shattuck}}]{jamming:BoromandPRL2018}%
  \BibitemOpen
  \bibfield  {author} {\bibinfo {author} {\bibfnamefont {A.}~\bibnamefont
  {Boromand}}, \bibinfo {author} {\bibfnamefont {A.}~\bibnamefont
  {Signoriello}}, \bibinfo {author} {\bibfnamefont {F.}~\bibnamefont {Ye}},
  \bibinfo {author} {\bibfnamefont {C.~S.}\ \bibnamefont {O'Hern}}, \ and\
  \bibinfo {author} {\bibfnamefont {M.~D.}\ \bibnamefont {Shattuck}},\
  }\href@noop {} {\bibfield  {journal} {\bibinfo  {journal} {Phys. Rev. Lett.}\
  }\textbf {\bibinfo {volume} {121}},\ \bibinfo {pages} {248003} (\bibinfo
  {year} {2018})}\BibitemShut {NoStop}%
\bibitem [{\citenamefont {Treado}\ \emph {et~al.}(2021)\citenamefont {Treado},
  \citenamefont {Wang}, \citenamefont {Boromand}, \citenamefont {Murrell},
  \citenamefont {Shattuck},\ and\ \citenamefont
  {O'Hern}}]{jamming:TreadoPRM2021}%
  \BibitemOpen
  \bibfield  {author} {\bibinfo {author} {\bibfnamefont {J.~D.}\ \bibnamefont
  {Treado}}, \bibinfo {author} {\bibfnamefont {D.}~\bibnamefont {Wang}},
  \bibinfo {author} {\bibfnamefont {A.}~\bibnamefont {Boromand}}, \bibinfo
  {author} {\bibfnamefont {M.~P.}\ \bibnamefont {Murrell}}, \bibinfo {author}
  {\bibfnamefont {M.~D.}\ \bibnamefont {Shattuck}}, \ and\ \bibinfo {author}
  {\bibfnamefont {C.~S.}\ \bibnamefont {O'Hern}},\ }\href@noop {} {\bibfield
  {journal} {\bibinfo  {journal} {Phys. Rev. Materials}\ }\textbf {\bibinfo
  {volume} {5}},\ \bibinfo {pages} {055605} (\bibinfo {year}
  {2021})}\BibitemShut {NoStop}%
\bibitem [{\citenamefont {Niklas}\ and\ \citenamefont
  {Spatz}(2012)}]{plants:NiklasUChicago2012}%
  \BibitemOpen
  \bibfield  {author} {\bibinfo {author} {\bibfnamefont {K.~J.}\ \bibnamefont
  {Niklas}}\ and\ \bibinfo {author} {\bibfnamefont {H.-C.}\ \bibnamefont
  {Spatz}},\ }\href@noop {} {\emph {\bibinfo {title} {Plant Physics}}}\
  (\bibinfo  {publisher} {University of Chicago Press},\ \bibinfo {year}
  {2012})\BibitemShut {NoStop}%
\bibitem [{\citenamefont {Lionetti}\ \emph {et~al.}(2015)\citenamefont
  {Lionetti}, \citenamefont {Cervone},\ and\ \citenamefont {{De
  Lorenzo}}}]{physio:LionettiPhytochem2015}%
  \BibitemOpen
  \bibfield  {author} {\bibinfo {author} {\bibfnamefont {V.}~\bibnamefont
  {Lionetti}}, \bibinfo {author} {\bibfnamefont {F.}~\bibnamefont {Cervone}}, \
  and\ \bibinfo {author} {\bibfnamefont {G.}~\bibnamefont {{De Lorenzo}}},\
  }\href@noop {} {\bibfield  {journal} {\bibinfo  {journal} {Phytochemistry}\
  }\textbf {\bibinfo {volume} {112}},\ \bibinfo {pages} {188} (\bibinfo {year}
  {2015})}\BibitemShut {NoStop}%
\bibitem [{\citenamefont {Sifton}(1945)}]{plants:SiftonBotRev1945}%
  \BibitemOpen
  \bibfield  {author} {\bibinfo {author} {\bibfnamefont {H.~B.}\ \bibnamefont
  {Sifton}},\ }\href@noop {} {\bibfield  {journal} {\bibinfo  {journal} {The
  Botanical Review}\ }\textbf {\bibinfo {volume} {11}},\ \bibinfo {pages} {108}
  (\bibinfo {year} {1945})}\BibitemShut {NoStop}%
\bibitem [{\citenamefont {Gao}\ \emph {et~al.}(2006)\citenamefont {Gao},
  \citenamefont {B\l{}awzdziewicz},\ and\ \citenamefont
  {O'Hern}}]{jamming:GaoPRE2006}%
  \BibitemOpen
  \bibfield  {author} {\bibinfo {author} {\bibfnamefont {G.-J.}\ \bibnamefont
  {Gao}}, \bibinfo {author} {\bibfnamefont {J.}~\bibnamefont
  {B\l{}awzdziewicz}}, \ and\ \bibinfo {author} {\bibfnamefont {C.~S.}\
  \bibnamefont {O'Hern}},\ }\href@noop {} {\bibfield  {journal} {\bibinfo
  {journal} {Phys. Rev. E}\ }\textbf {\bibinfo {volume} {74}} (\bibinfo {year}
  {2006})}\BibitemShut {NoStop}%
\bibitem [{\citenamefont {Rhizopoulou}\ and\ \citenamefont
  {Psaras}(2003)}]{plants:RhizopoulouAnnBot2003}%
  \BibitemOpen
  \bibfield  {author} {\bibinfo {author} {\bibfnamefont {S.}~\bibnamefont
  {Rhizopoulou}}\ and\ \bibinfo {author} {\bibfnamefont {G.~K.}\ \bibnamefont
  {Psaras}},\ }\href@noop {} {\bibfield  {journal} {\bibinfo  {journal} {Annals
  of Botany}\ }\textbf {\bibinfo {volume} {92}},\ \bibinfo {pages} {377}
  (\bibinfo {year} {2003})}\BibitemShut {NoStop}%
\bibitem [{\citenamefont {Harwood}\ \emph {et~al.}(2020)\citenamefont
  {Harwood}, \citenamefont {Goodman}, \citenamefont {Gudmundsdottir},
  \citenamefont {Huynh}, \citenamefont {Musulin}, \citenamefont {Song},\ and\
  \citenamefont {Barbour}}]{plants:HarwoodNewPhyto2020}%
  \BibitemOpen
  \bibfield  {author} {\bibinfo {author} {\bibfnamefont {R.}~\bibnamefont
  {Harwood}}, \bibinfo {author} {\bibfnamefont {E.}~\bibnamefont {Goodman}},
  \bibinfo {author} {\bibfnamefont {M.}~\bibnamefont {Gudmundsdottir}},
  \bibinfo {author} {\bibfnamefont {M.}~\bibnamefont {Huynh}}, \bibinfo
  {author} {\bibfnamefont {Q.}~\bibnamefont {Musulin}}, \bibinfo {author}
  {\bibfnamefont {M.}~\bibnamefont {Song}}, \ and\ \bibinfo {author}
  {\bibfnamefont {M.~M.}\ \bibnamefont {Barbour}},\ }\href@noop {} {\bibfield
  {journal} {\bibinfo  {journal} {New Phytologist}\ }\textbf {\bibinfo {volume}
  {225}},\ \bibinfo {pages} {2567} (\bibinfo {year} {2020})}\BibitemShut
  {NoStop}%
\bibitem [{\citenamefont {Goriely}(2017)}]{growth:GorielySpringer2017}%
  \BibitemOpen
  \bibfield  {author} {\bibinfo {author} {\bibfnamefont {A.}~\bibnamefont
  {Goriely}},\ }\enquote {\bibinfo {title} {Growing on a line},}\ in\
  \href@noop {} {\emph {\bibinfo {booktitle} {The Mathematics and Mechanics of
  Biological Growth}}}\ (\bibinfo  {publisher} {Springer New York},\ \bibinfo
  {year} {2017})\ pp.\ \bibinfo {pages} {63--96}\BibitemShut {NoStop}%
\bibitem [{\citenamefont {Zonia}\ \emph {et~al.}(2006)\citenamefont {Zonia},
  \citenamefont {M{\"u}ller},\ and\ \citenamefont
  {Munnik}}]{plants:ZoniaCellBiochemBiophys2006}%
  \BibitemOpen
  \bibfield  {author} {\bibinfo {author} {\bibfnamefont {L.}~\bibnamefont
  {Zonia}}, \bibinfo {author} {\bibfnamefont {M.}~\bibnamefont {M{\"u}ller}}, \
  and\ \bibinfo {author} {\bibfnamefont {T.}~\bibnamefont {Munnik}},\
  }\href@noop {} {\bibfield  {journal} {\bibinfo  {journal} {Cell Biochemistry
  and Biophysics}\ }\textbf {\bibinfo {volume} {46}},\ \bibinfo {pages} {209}
  (\bibinfo {year} {2006})}\BibitemShut {NoStop}%
\bibitem [{\citenamefont {Kroeger}\ \emph {et~al.}(2011)\citenamefont
  {Kroeger}, \citenamefont {Zerzour},\ and\ \citenamefont
  {Geitmann}}]{plants:KroegerPLOSONE2011}%
  \BibitemOpen
  \bibfield  {author} {\bibinfo {author} {\bibfnamefont {J.~H.}\ \bibnamefont
  {Kroeger}}, \bibinfo {author} {\bibfnamefont {R.}~\bibnamefont {Zerzour}}, \
  and\ \bibinfo {author} {\bibfnamefont {A.}~\bibnamefont {Geitmann}},\
  }\href@noop {} {\bibfield  {journal} {\bibinfo  {journal} {PLOS ONE}\
  }\textbf {\bibinfo {volume} {6}},\ \bibinfo {pages} {1} (\bibinfo {year}
  {2011})}\BibitemShut {NoStop}%
\bibitem [{\citenamefont {Jacobs}\ and\ \citenamefont
  {Thorpe}(1996)}]{pebble:JacobsPRE1996}%
  \BibitemOpen
  \bibfield  {author} {\bibinfo {author} {\bibfnamefont {D.~J.}\ \bibnamefont
  {Jacobs}}\ and\ \bibinfo {author} {\bibfnamefont {M.~F.}\ \bibnamefont
  {Thorpe}},\ }\href@noop {} {\bibfield  {journal} {\bibinfo  {journal} {Phys.
  Rev. E}\ }\textbf {\bibinfo {volume} {53}} (\bibinfo {year}
  {1996})}\BibitemShut {NoStop}%
\bibitem [{\citenamefont {Donev}\ \emph {et~al.}(2007)\citenamefont {Donev},
  \citenamefont {Connelly}, \citenamefont {Stillinger},\ and\ \citenamefont
  {Torquato}}]{ellipse:DonevPRE2007}%
  \BibitemOpen
  \bibfield  {author} {\bibinfo {author} {\bibfnamefont {A.}~\bibnamefont
  {Donev}}, \bibinfo {author} {\bibfnamefont {R.}~\bibnamefont {Connelly}},
  \bibinfo {author} {\bibfnamefont {F.~H.}\ \bibnamefont {Stillinger}}, \ and\
  \bibinfo {author} {\bibfnamefont {S.}~\bibnamefont {Torquato}},\ }\href@noop
  {} {\bibfield  {journal} {\bibinfo  {journal} {Phys. Rev. E}\ }\textbf
  {\bibinfo {volume} {75}} (\bibinfo {year} {2007})}\BibitemShut {NoStop}%
\bibitem [{\citenamefont {Frensch}\ and\ \citenamefont
  {Schulze}(1988)}]{plants:FrenschPlanta1988}%
  \BibitemOpen
  \bibfield  {author} {\bibinfo {author} {\bibfnamefont {J.}~\bibnamefont
  {Frensch}}\ and\ \bibinfo {author} {\bibfnamefont {E.-D.}\ \bibnamefont
  {Schulze}},\ }\href@noop {} {\bibfield  {journal} {\bibinfo  {journal}
  {Planta}\ }\textbf {\bibinfo {volume} {173}},\ \bibinfo {pages} {554}
  (\bibinfo {year} {1988})}\BibitemShut {NoStop}%
\bibitem [{\citenamefont {Nonami}\ and\ \citenamefont
  {Schulze}(1989)}]{plants:NonamiPlanta1989}%
  \BibitemOpen
  \bibfield  {author} {\bibinfo {author} {\bibfnamefont {H.}~\bibnamefont
  {Nonami}}\ and\ \bibinfo {author} {\bibfnamefont {E.~D.}\ \bibnamefont
  {Schulze}},\ }\href@noop {} {\bibfield  {journal} {\bibinfo  {journal}
  {Planta}\ }\textbf {\bibinfo {volume} {177}},\ \bibinfo {pages} {35}
  (\bibinfo {year} {1989})}\BibitemShut {NoStop}%
\bibitem [{\citenamefont {Th{\"u}rmer}\ \emph {et~al.}(1999)\citenamefont
  {Th{\"u}rmer}, \citenamefont {Zhu}, \citenamefont {Gierlinger}, \citenamefont
  {Schneider}, \citenamefont {Benkert}, \citenamefont {Ge{\ss}ner},
  \citenamefont {Herrmann}, \citenamefont {Bentrup},\ and\ \citenamefont
  {Zimmerniann}}]{plants:ThurmerProtoplasma1999}%
  \BibitemOpen
  \bibfield  {author} {\bibinfo {author} {\bibfnamefont {F.}~\bibnamefont
  {Th{\"u}rmer}}, \bibinfo {author} {\bibfnamefont {J.~J.}\ \bibnamefont
  {Zhu}}, \bibinfo {author} {\bibfnamefont {N.}~\bibnamefont {Gierlinger}},
  \bibinfo {author} {\bibfnamefont {H.}~\bibnamefont {Schneider}}, \bibinfo
  {author} {\bibfnamefont {R.}~\bibnamefont {Benkert}}, \bibinfo {author}
  {\bibfnamefont {P.}~\bibnamefont {Ge{\ss}ner}}, \bibinfo {author}
  {\bibfnamefont {B.}~\bibnamefont {Herrmann}}, \bibinfo {author}
  {\bibfnamefont {F.~W.}\ \bibnamefont {Bentrup}}, \ and\ \bibinfo {author}
  {\bibfnamefont {U.}~\bibnamefont {Zimmerniann}},\ }\href@noop {} {\bibfield
  {journal} {\bibinfo  {journal} {Protoplasma}\ }\textbf {\bibinfo {volume}
  {206}},\ \bibinfo {pages} {152} (\bibinfo {year} {1999})}\BibitemShut
  {NoStop}%
\bibitem [{\citenamefont {Roddy}\ \emph {et~al.}(2019)\citenamefont {Roddy},
  \citenamefont {Jiang}, \citenamefont {Cao}, \citenamefont {Simonin},\ and\
  \citenamefont {Brodersen}}]{physio:RoddyNewPhyto2019}%
  \BibitemOpen
  \bibfield  {author} {\bibinfo {author} {\bibfnamefont {A.~B.}\ \bibnamefont
  {Roddy}}, \bibinfo {author} {\bibfnamefont {G.-F.}\ \bibnamefont {Jiang}},
  \bibinfo {author} {\bibfnamefont {K.}~\bibnamefont {Cao}}, \bibinfo {author}
  {\bibfnamefont {K.~A.}\ \bibnamefont {Simonin}}, \ and\ \bibinfo {author}
  {\bibfnamefont {C.~R.}\ \bibnamefont {Brodersen}},\ }\href@noop {} {\bibfield
   {journal} {\bibinfo  {journal} {New Phytologist}\ }\textbf {\bibinfo
  {volume} {223}},\ \bibinfo {pages} {193} (\bibinfo {year}
  {2019})}\BibitemShut {NoStop}%
\bibitem [{\citenamefont {Kutschera}\ and\ \citenamefont
  {Niklas}(2007)}]{plants:KutscheraJPlantPhys2007}%
  \BibitemOpen
  \bibfield  {author} {\bibinfo {author} {\bibfnamefont {U.}~\bibnamefont
  {Kutschera}}\ and\ \bibinfo {author} {\bibfnamefont {K.}~\bibnamefont
  {Niklas}},\ }\href@noop {} {\bibfield  {journal} {\bibinfo  {journal}
  {Journal of Plant Physiology}\ }\textbf {\bibinfo {volume} {164}},\ \bibinfo
  {pages} {1395} (\bibinfo {year} {2007})}\BibitemShut {NoStop}%
\bibitem [{\citenamefont {Abe}\ \emph {et~al.}(2003)\citenamefont {Abe},
  \citenamefont {Katsumata}, \citenamefont {Komeda},\ and\ \citenamefont
  {Takahashi}}]{plants:AbeDev2003}%
  \BibitemOpen
  \bibfield  {author} {\bibinfo {author} {\bibfnamefont {M.}~\bibnamefont
  {Abe}}, \bibinfo {author} {\bibfnamefont {H.}~\bibnamefont {Katsumata}},
  \bibinfo {author} {\bibfnamefont {Y.}~\bibnamefont {Komeda}}, \ and\ \bibinfo
  {author} {\bibfnamefont {T.}~\bibnamefont {Takahashi}},\ }\href@noop {}
  {\bibfield  {journal} {\bibinfo  {journal} {Development}\ }\textbf {\bibinfo
  {volume} {130}},\ \bibinfo {pages} {635} (\bibinfo {year}
  {2003})}\BibitemShut {NoStop}%
\bibitem [{\citenamefont {Panteris}\ \emph {et~al.}(1993)\citenamefont
  {Panteris}, \citenamefont {Apostolakos},\ and\ \citenamefont
  {Galatis}}]{plants:PanterisProtoplasma1993}%
  \BibitemOpen
  \bibfield  {author} {\bibinfo {author} {\bibfnamefont {E.}~\bibnamefont
  {Panteris}}, \bibinfo {author} {\bibfnamefont {P.}~\bibnamefont
  {Apostolakos}}, \ and\ \bibinfo {author} {\bibfnamefont {B.}~\bibnamefont
  {Galatis}},\ }\href@noop {} {\bibfield  {journal} {\bibinfo  {journal}
  {Protoplasma}\ }\textbf {\bibinfo {volume} {172}},\ \bibinfo {pages} {97}
  (\bibinfo {year} {1993})}\BibitemShut {NoStop}%
\bibitem [{\citenamefont {Bidhendi}\ and\ \citenamefont
  {Geitmann}(2015)}]{dev:BidhendiJExpBot2015}%
  \BibitemOpen
  \bibfield  {author} {\bibinfo {author} {\bibfnamefont {A.~J.}\ \bibnamefont
  {Bidhendi}}\ and\ \bibinfo {author} {\bibfnamefont {A.}~\bibnamefont
  {Geitmann}},\ }\href@noop {} {\bibfield  {journal} {\bibinfo  {journal}
  {Journal of Experimental Botany}\ }\textbf {\bibinfo {volume} {67}},\
  \bibinfo {pages} {449} (\bibinfo {year} {2015})}\BibitemShut {NoStop}%
\bibitem [{\citenamefont {Wang}\ \emph {et~al.}(2021)\citenamefont {Wang},
  \citenamefont {Treado}, \citenamefont {Boromand}, \citenamefont {Norwick},
  \citenamefont {Murrell}, \citenamefont {Shattuck},\ and\ \citenamefont
  {O{'}Hern}}]{jamming:WangSM2021}%
  \BibitemOpen
  \bibfield  {author} {\bibinfo {author} {\bibfnamefont {D.}~\bibnamefont
  {Wang}}, \bibinfo {author} {\bibfnamefont {J.~D.}\ \bibnamefont {Treado}},
  \bibinfo {author} {\bibfnamefont {A.}~\bibnamefont {Boromand}}, \bibinfo
  {author} {\bibfnamefont {B.}~\bibnamefont {Norwick}}, \bibinfo {author}
  {\bibfnamefont {M.~P.}\ \bibnamefont {Murrell}}, \bibinfo {author}
  {\bibfnamefont {M.~D.}\ \bibnamefont {Shattuck}}, \ and\ \bibinfo {author}
  {\bibfnamefont {C.~S.}\ \bibnamefont {O{'}Hern}},\ }\href@noop {} {\bibfield
  {journal} {\bibinfo  {journal} {Soft Matter}\ }\textbf {\bibinfo {volume}
  {17}},\ \bibinfo {pages} {9901} (\bibinfo {year} {2021})}\BibitemShut
  {NoStop}%
\bibitem [{\citenamefont {Zhang}\ and\ \citenamefont
  {Ambrose}(2022)}]{plants:ZhangNatPlants2022}%
  \BibitemOpen
  \bibfield  {author} {\bibinfo {author} {\bibfnamefont {L.}~\bibnamefont
  {Zhang}}\ and\ \bibinfo {author} {\bibfnamefont {C.}~\bibnamefont
  {Ambrose}},\ }\href@noop {} {\bibfield  {journal} {\bibinfo  {journal}
  {Nature Plants}\ }\textbf {\bibinfo {volume} {8}},\ \bibinfo {pages} {682}
  (\bibinfo {year} {2022})}\BibitemShut {NoStop}%
\bibitem [{\citenamefont {Schindelin}\ \emph {et~al.}(2012)\citenamefont
  {Schindelin}, \citenamefont {Arganda-Carreras}, \citenamefont {Frise},
  \citenamefont {Kaynig}, \citenamefont {Longair}, \citenamefont {Pietzsch},
  \citenamefont {Preibisch}, \citenamefont {Rueden}, \citenamefont {Saalfeld},
  \citenamefont {Schmid}, \citenamefont {Tinevez}, \citenamefont {White},
  \citenamefont {Hartenstein}, \citenamefont {Eliceiri}, \citenamefont
  {Tomancak},\ and\ \citenamefont {Cardona}}]{fiji:SchindelinNatMethods2012}%
  \BibitemOpen
  \bibfield  {author} {\bibinfo {author} {\bibfnamefont {J.}~\bibnamefont
  {Schindelin}}, \bibinfo {author} {\bibfnamefont {I.}~\bibnamefont
  {Arganda-Carreras}}, \bibinfo {author} {\bibfnamefont {E.}~\bibnamefont
  {Frise}}, \bibinfo {author} {\bibfnamefont {V.}~\bibnamefont {Kaynig}},
  \bibinfo {author} {\bibfnamefont {M.}~\bibnamefont {Longair}}, \bibinfo
  {author} {\bibfnamefont {T.}~\bibnamefont {Pietzsch}}, \bibinfo {author}
  {\bibfnamefont {S.}~\bibnamefont {Preibisch}}, \bibinfo {author}
  {\bibfnamefont {C.}~\bibnamefont {Rueden}}, \bibinfo {author} {\bibfnamefont
  {S.}~\bibnamefont {Saalfeld}}, \bibinfo {author} {\bibfnamefont
  {B.}~\bibnamefont {Schmid}}, \bibinfo {author} {\bibfnamefont {J.-Y.}\
  \bibnamefont {Tinevez}}, \bibinfo {author} {\bibfnamefont {D.~J.}\
  \bibnamefont {White}}, \bibinfo {author} {\bibfnamefont {V.}~\bibnamefont
  {Hartenstein}}, \bibinfo {author} {\bibfnamefont {K.}~\bibnamefont
  {Eliceiri}}, \bibinfo {author} {\bibfnamefont {P.}~\bibnamefont {Tomancak}},
  \ and\ \bibinfo {author} {\bibfnamefont {A.}~\bibnamefont {Cardona}},\
  }\href@noop {} {\bibfield  {journal} {\bibinfo  {journal} {Nature Methods}\
  }\textbf {\bibinfo {volume} {9}},\ \bibinfo {pages} {676} (\bibinfo {year}
  {2012})}\BibitemShut {NoStop}%
\bibitem [{\citenamefont {Wan}\ \emph {et~al.}(2012)\citenamefont {Wan},
  \citenamefont {Otsuna}, \citenamefont {Chien},\ and\ \citenamefont
  {Hansen}}]{fluo:WanIEEE2012}%
  \BibitemOpen
  \bibfield  {author} {\bibinfo {author} {\bibfnamefont {Y.}~\bibnamefont
  {Wan}}, \bibinfo {author} {\bibfnamefont {H.}~\bibnamefont {Otsuna}},
  \bibinfo {author} {\bibfnamefont {C.-B.}\ \bibnamefont {Chien}}, \ and\
  \bibinfo {author} {\bibfnamefont {C.}~\bibnamefont {Hansen}},\ }in\
  \href@noop {} {\emph {\bibinfo {booktitle} {2012 IEEE Pacific Visualization
  Symposium}}}\ (\bibinfo {year} {2012})\ pp.\ \bibinfo {pages}
  {201--208}\BibitemShut {NoStop}%
\bibitem [{\citenamefont {Paganin}\ \emph {et~al.}(2002)\citenamefont
  {Paganin}, \citenamefont {Mayo}, \citenamefont {Gureyev}, \citenamefont
  {Miller},\ and\ \citenamefont {Wilkins}}]{microct:paganinJMicro2002}%
  \BibitemOpen
  \bibfield  {author} {\bibinfo {author} {\bibfnamefont {D.}~\bibnamefont
  {Paganin}}, \bibinfo {author} {\bibfnamefont {S.~C.}\ \bibnamefont {Mayo}},
  \bibinfo {author} {\bibfnamefont {T.~E.}\ \bibnamefont {Gureyev}}, \bibinfo
  {author} {\bibfnamefont {P.~R.}\ \bibnamefont {Miller}}, \ and\ \bibinfo
  {author} {\bibfnamefont {S.~W.}\ \bibnamefont {Wilkins}},\ }\href@noop {}
  {\bibfield  {journal} {\bibinfo  {journal} {Journal of Microscopy}\ }\textbf
  {\bibinfo {volume} {206}},\ \bibinfo {pages} {33} (\bibinfo {year}
  {2002})}\BibitemShut {NoStop}%
\bibitem [{\citenamefont {Th{\'e}roux-Rancourt}\ \emph
  {et~al.}(2017)\citenamefont {Th{\'e}roux-Rancourt}, \citenamefont {Earles},
  \citenamefont {Gilbert}, \citenamefont {Zwieniecki}, \citenamefont {Boyce},
  \citenamefont {McElrone},\ and\ \citenamefont
  {Brodersen}}]{physio:TherouxRancourtNewPhyto2017}%
  \BibitemOpen
  \bibfield  {author} {\bibinfo {author} {\bibfnamefont {G.}~\bibnamefont
  {Th{\'e}roux-Rancourt}}, \bibinfo {author} {\bibfnamefont {J.~M.}\
  \bibnamefont {Earles}}, \bibinfo {author} {\bibfnamefont {M.~E.}\
  \bibnamefont {Gilbert}}, \bibinfo {author} {\bibfnamefont {M.~A.}\
  \bibnamefont {Zwieniecki}}, \bibinfo {author} {\bibfnamefont {C.~K.}\
  \bibnamefont {Boyce}}, \bibinfo {author} {\bibfnamefont {A.~J.}\ \bibnamefont
  {McElrone}}, \ and\ \bibinfo {author} {\bibfnamefont {C.~R.}\ \bibnamefont
  {Brodersen}},\ }\href@noop {} {\bibfield  {journal} {\bibinfo  {journal} {New
  Phytologist}\ }\textbf {\bibinfo {volume} {215}},\ \bibinfo {pages} {1609}
  (\bibinfo {year} {2017})}\BibitemShut {NoStop}%
\bibitem [{\citenamefont {Vincent}\ and\ \citenamefont
  {Soille}(1991)}]{watershed:VincentIEEETransPatt1991}%
  \BibitemOpen
  \bibfield  {author} {\bibinfo {author} {\bibfnamefont {L.}~\bibnamefont
  {Vincent}}\ and\ \bibinfo {author} {\bibfnamefont {P.}~\bibnamefont
  {Soille}},\ }\href@noop {} {\bibfield  {journal} {\bibinfo  {journal} {IEEE
  Transactions on Pattern Analysis and Machine Intelligence}\ }\textbf
  {\bibinfo {volume} {13}},\ \bibinfo {pages} {583} (\bibinfo {year}
  {1991})}\BibitemShut {NoStop}%
\end{thebibliography}%


\begin{thebibliography}{8}
\providecommand{\natexlab}[1]{#1}
\providecommand{\url}[1]{\texttt{#1}}
\expandafter\ifx\csname urlstyle\endcsname\relax
  \providecommand{\doi}[1]{doi: #1}\else
  \providecommand{\doi}{doi: \begingroup \urlstyle{rm}\Url}\fi

\bibitem[Allen and Tildesley(2017)]{sim:AllenOxford2017}
M.~P. Allen and D.~J. Tildesley.
\newblock \emph{Computer Simulation of Liquids}.
\newblock Oxford University Press, 2nd edition, 2017.

\bibitem[Andersen(1980)]{constp}
H.~C. Andersen.
\newblock Molecular dynamics simulations at constant pressure and/or
  temperature.
\newblock \emph{The Journal of Chemical Physics}, 72\penalty0 (4):\penalty0
  2384--2393, 1980.

\bibitem[Bitzek et~al.(2006)Bitzek, Koskinen, G\"ahler, Moseler, and
  Gumbsch]{fire}
E.~Bitzek, P.~Koskinen, F.~G\"ahler, M.~Moseler, and P.~Gumbsch.
\newblock Structural relaxation made simple.
\newblock \emph{Phys. Rev. Lett.}, 97:\penalty0 170201, Oct 2006.
\newblock \doi{10.1103/PhysRevLett.97.170201}.

\bibitem[Daher and Braybrook(2015)]{physio:DaherFrontInPlantSci2015}
F.~B. Daher and S.~A. Braybrook.
\newblock How to let go: pectin and plant cell adhesion.
\newblock \emph{Frontiers in Plant Science}, 6:\penalty0 523, 2015.

\bibitem[Gu{\'e}nol{\'e} et~al.(2020)Gu{\'e}nol{\'e}, N{\"o}hring, Vaid,
  Houll{\'e}, Xie, Prakash, and Bitzek]{fire2}
J.~Gu{\'e}nol{\'e}, W.~G. N{\"o}hring, A.~Vaid, F.~Houll{\'e}, Z.~Xie,
  A.~Prakash, and E.~Bitzek.
\newblock Assessment and optimization of the fast inertial relaxation engine
  (fire) for energy minimization in atomistic simulations and its
  implementation in lammps.
\newblock \emph{Computational Materials Science}, 175:\penalty0 109584, 2020.

\bibitem[Lionetti et~al.(2015)Lionetti, Cervone, and {De
  Lorenzo}]{physio:LionettiPhytochem2015}
V.~Lionetti, F.~Cervone, and G.~{De Lorenzo}.
\newblock A lower content of de-methylesterified homogalacturonan improves
  enzymatic cell separation and isolation of mesophyll protoplasts in
  arabidopsis.
\newblock \emph{Phytochemistry}, 112:\penalty0 188--194, 2015.

\bibitem[Seago et~al.(2005)Seago, Marsh, Stevens, Soukup, Votrubov{\'a}, and
  Enstone]{physio:SeagoAnnBot2005}
J.~Seago, James~L., L.~C. Marsh, K.~J. Stevens, A.~Soukup, O.~Votrubov{\'a},
  and D.~E. Enstone.
\newblock A re-examination of the root cortex in wetland flowering plants with
  respect to aerenchyma.
\newblock \emph{Annals of Botany}, 96\penalty0 (4):\penalty0 565--579, 08 2005.

\bibitem[Zhang et~al.(2021)Zhang, McEvoy, Le, and
  Ambrose]{physio:ZhangPlantCell2021}
L.~Zhang, D.~McEvoy, Y.~Le, and C.~Ambrose.
\newblock {Live imaging of microtubule organization, cell expansion, and
  intercellular space formation in Arabidopsis leaf spongy mesophyll cells}.
\newblock \emph{The Plant Cell}, 33\penalty0 (3):\penalty0 623--641, 12 2021.

\end{thebibliography}

\end{document}


\maketitle
\tableofcontents

\beginsupplement

\section{Deformable polygon model for spongy mesophyll cells}\label{app:dpm}

In this section, we derive the analytical forms of the forces on each vertex $k$ on deformable polygon $\mu$. In general, vertices are acted upon by shape forces (i.e. the area, perimeter, and bending terms in Eq. $1$ in the main text), and by interaction forces (i.e. the intercellular potentials in Eq. $2$ in the main text). We denote the potential energy corresponding to each force by $U_{\rm a}$, $U_{\rm l}$, $U_{\rm b}$, and $U_{\rm int}$, respectively. The force on vertex $k$ on cell $\mu$ is the vector gradient of all potential energy terms with respect to the coordinate $\vec{r}_{k\mu} = x_{k\mu}\hat{e}_x + y_{k\mu} \hat{e}_y$, i.e.
\begin{equation}
    \vec{F}_{k\mu} = -\pdv{U}{\vec{r}_{k\mu}} \equiv -\pdv{U}{x_{k\mu}}\hat{e}_x - \pdv{U}{y_{k\mu}}\hat{e}_y.
\end{equation}

\subsection{Area force}
The force on vertex $k$ of cell $\mu$ due to deviations in the preferred area is given by $-\pdv*{U_{\rm a}}{\vec{r}_{k\mu}}$. For convenience, we will neglect the index $\mu$ when describing purely shape forces, as only vertices on cell $\mu$ are involved. The area force is defined as
\begin{equation}
    \vec{F}_k^{\rm a} = -\pdv{U_{\rm a}}{\vec{r}_k} = -\frac{\epsilon_a}{a_0}\qty(\frac{a}{a_0}-1)\pdv{a}{\vec{r}_k}.
\end{equation}
The polygonal area $a$ is obtained by applying Gauss's ``shoelace" formula,
\begin{equation}
    a = \frac{1}{2}\sum_{k=1}^n x_ky_{k+1} - x_{k+1}y_k,
\end{equation}
where we take the index $n+1$ in the summation to be $1$, and the index $0$ to be $n$. Therefore, the derivatives in the $x$ and $y$ directions are not symmetric, rather $\pdv*{a}{x_k} = (y_{k+1}-y_{k-1})/2$ and $\pdv*{a}{y_k} = (x_{k-1}-x_{k+1})/2$. Therefore, we find
\begin{equation}
    \vec{F}_k^{\rm a} = \frac{\epsilon_a}{2a_0}\qty(\frac{a}{a_0}-1)\qty[(y_{k-1}-y_{k+1})\hat{e}_x + (x_{k+1}-x_{k-1})\hat{e}_y]. 
\end{equation}

\begin{figure}[!t]
    \centering
    \includegraphics[width=0.75\textwidth]{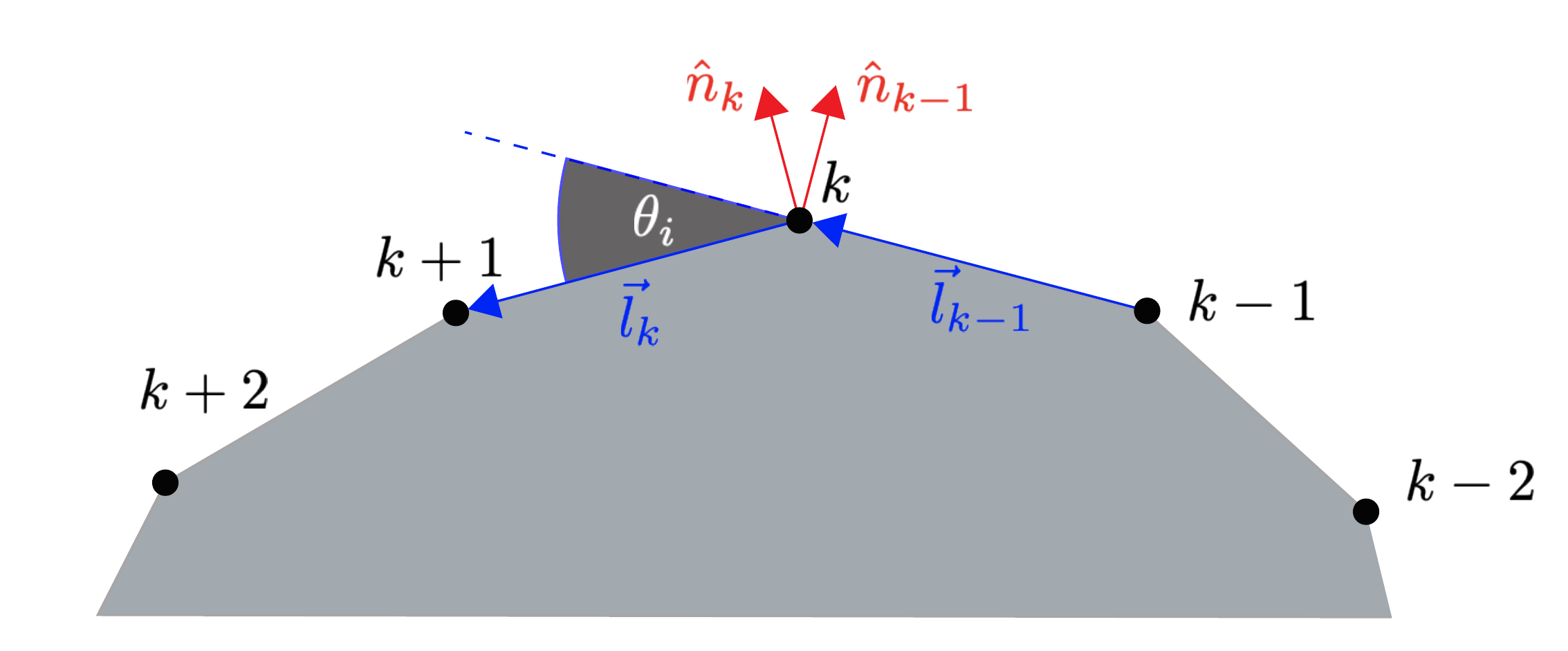}
    \caption{Geometry of regions near individual vertices $\ldots, k-1, k, k+1,\ldots$ on DP model cells. Segment vectors $\vec{l}_k = \vec{r}_{k+1} - \vec{r}_k$ drawn in blue define segment lengths $l_k = \abs{\vec{l}_k}$, and the bending angle $\theta_k$ is the angle between $\vec{l}_{k-1}$ and $\vec{l}_k$. The unit vectors $\hat{n}_k$ and $\hat{n}_{k-1}$ drawn in red are normal to $\vec{l}_k$ and $\vec{l}_{k-1}$, respectively.}
    \label{appfig:angles}
\end{figure}

\subsection{Perimeter force}
The force on vertex $k$ due to deviations in the segment length $l_k$ (see Fig.~\ref{appfig:angles}) from its preferred value is
\begin{equation}
    \vec{F}_k^{\rm l} = -\pdv{U_{\rm l}}{\vec{r}_k} = -\epsilon_l\qty[\frac{1}{l_{0k}}\qty(\frac{l_k}{l_{0k}} - 1)\pdv{l_k}{\vec{r}_k} + \frac{1}{l_{0k-1}}\qty(\frac{l_{k-1}}{l_{0k-1}} - 1)\pdv{l_{k-1}}{\vec{r}_k}].
\end{equation}
Derivatives of $l_k$ with respect to $\vec{r}_k$ \emph{and} $\vec{r}_{k+1}$ are non-zero. Taking these derivatives with respect to $\vec{r}_k$, we obtain $\pdv*{l_k}{\vec{r}_k} = -\hat{l}_k$ and $\pdv*{l_{k-1}}{\vec{r}_k} = \hat{l}_{k-1}$, where $\hat{l}_k \equiv \vec{l}_k/l_k$. Therefore, the perimeter force on vertex $k$ is
\begin{equation}
    \vec{F}_k^{\rm l} = \epsilon_l\qty[\qty(\frac{l_k}{l_{0k}} - 1)\frac{\hat{l}_k}{l_{0k}} - \qty(\frac{l_{k-1}}{l_{0k-1}} - 1)\frac{\hat{l}_{k-1}}{l_{0k-1}}].
\end{equation}

\subsection{Bending force}
The force on vertex $k$ due to deviations in the local bending angle $\theta_k$ from its preferred value, $\delta\theta_k = \theta_k - \theta_{0k}$, is given by 
\begin{equation}\label{eq:bforce_1}
    \vec{F}_k^{\rm b} = -\pdv{U_{\rm b}}{\vec{r}_k} = -\epsilon_b\qty[\delta\theta_{k-1}\pdv{\theta_{k-1}}{\vec{r}_k} + \delta\theta_k\pdv{\theta_k}{\vec{r}_k} + \delta\theta_{k+1}\pdv{\theta_{k+1}}{\vec{r}_k}].
\end{equation}
As shown in Fig.~\ref{appfig:angles}, we define the local bending angle at vertex $k$ to be the angle between the vectors $\vec{l}_{k-1}$ and $\vec{l}_k$. This angle can be calculated using the definitions $\sin\theta_k = \abs{\hat{l}_{k-1}\times\hat{l}_k}$, $\cos\theta_k = \hat{l}_k\cdot\hat{l}_{k-1}$, and $\theta_k = \tan^{-1}\qty(\sin\theta_k/\cos\theta_k)$, i.e.
\begin{equation}
    \theta_k = \tan^{-1}\qty(\frac{l_{k-1,x}l_{k,y} - l_{k,x}l_{k-1,y}}{l_{k-1,x}l_{k-1,x} + l_{k,x}l_{k,y}}),
\end{equation}
where $l_{k,\xi}$ is the $\xi=x$- or $y$-component of the segment vector $\vec{l}_k$. We define the temporary variables
\begin{equation}\label{eq:skck}
    s_k = l_{k-1,x}l_{k,y} - l_{k,x}l_{k-1,y}~{\rm and}~c_k = l_{k-1,x}l_{k-1,x} + l_{k,x}l_{k,y},
\end{equation}
which gives 
\begin{equation}\label{eq:thetak_1}
    \pdv{\theta_k}{\vec{r}_k} = \frac{c_k\pdv{s_k}{\vec{r}_k} - s_k\pdv{c_k}{\vec{r}_k}}{c_k^2 + s_k^2},
\end{equation}
where we have used $\pdv*{\tan^{-1}(x)}{x} = (1 + x^2)^{-1}$. The derivatives in the numerator of Eq.~\eqref{eq:thetak_1} are
\begin{equation}\label{eq:skckderivs}
\begin{split}
    \pdv{s_k}{\vec{r}_k} &= \qty(l_{k,y} + l_{k-1,y})\hat{e}_x - \qty(l_{k,x} + l_{k-1,x})\hat{e}_y\\
    \pdv{c_k}{\vec{r}_k} &= \qty(l_{k,y} - l_{k-1,y})\hat{e}_x + \qty(l_{k,y} - l_{k-1,y})\hat{e}_y, 
\end{split}
\end{equation}
and we note that $c_k^2 + s_k^2 = l_k^2 l_{k-1}^2$, given the definition of $c_k$ and $s_k$ in Eq.~\eqref{eq:skck}. Incorporating Eq.~\eqref{eq:skckderivs} into Eq.~\eqref{eq:thetak_1} yields
\begin{equation}\label{eq:thetak_2}
    \pdv{\theta_k}{\vec{r}_k} = \frac{\hat{n}_k}{l_k} + \frac{\hat{n}_{k-1}}{l_{k-1}},
\end{equation}
where $\vec{n}_k = l_{k,y}\hat{e}_x - l_{k,x}\hat{e}_y$ is the vector normal to $\vec{l}_k$, and is drawn in red as the unit vector $\hat{n}_k = \vec{n}_k/l_k$ in Fig.~\ref{appfig:angles}.

Because each polygonal cell forms a closed loop, $\sum_{i=1}^n \theta_i = 2\pi$, and thus $\theta_k = 2\pi - \sum_{i\neq k}\theta_i$. Therefore,
\begin{equation}
    \pdv{\theta_k}{\vec{r}_k} = -\pdv{\theta_{k-1}}{\vec{r}_k} - \pdv{\theta_{k+1}}{\vec{r}_k},
\end{equation}
as $\theta_k$ only depends on $\vec{r}_{k-1}$, $\vec{r}_k$, and $\vec{r}_{k+1}$. Comparing coefficients with Eq.~\eqref{eq:thetak_2}, we then obtain the derivatives of the other two angles with respect to $\vec{r}_k$:
\begin{equation}
    \pdv{\theta_{k-1}}{\vec{r}_k} = -\frac{\hat{n}_{k-1}}{l_{k-1}}~{\rm and}~ \pdv{\theta_{k+1}}{\vec{r}_k} = -\frac{\hat{n}_k}{l_k}.
\end{equation}
We can substitute the forms of the derivatives of each angle with respect to $\vec{r}_k$ into Eq.~\eqref{eq:bforce_1} to find
\begin{equation}
    \vec{F}_k^{\rm b} = \epsilon_b\qty[\qty(\frac{\delta\theta_{k-1} - \delta \theta_k}{l_{k-1}})\hat{n}_{k-1} + \qty(\frac{\delta\theta_{k+1}-\delta\theta_k}{l_k})\hat{n}_k].
\end{equation}
Bending forces therefore typically point normal to the cell surface.

\section{Bonds for cell-cell adhesion}\label{app:bonds}

\begin{figure}
    \centering
    \includegraphics[width=0.75\textwidth]{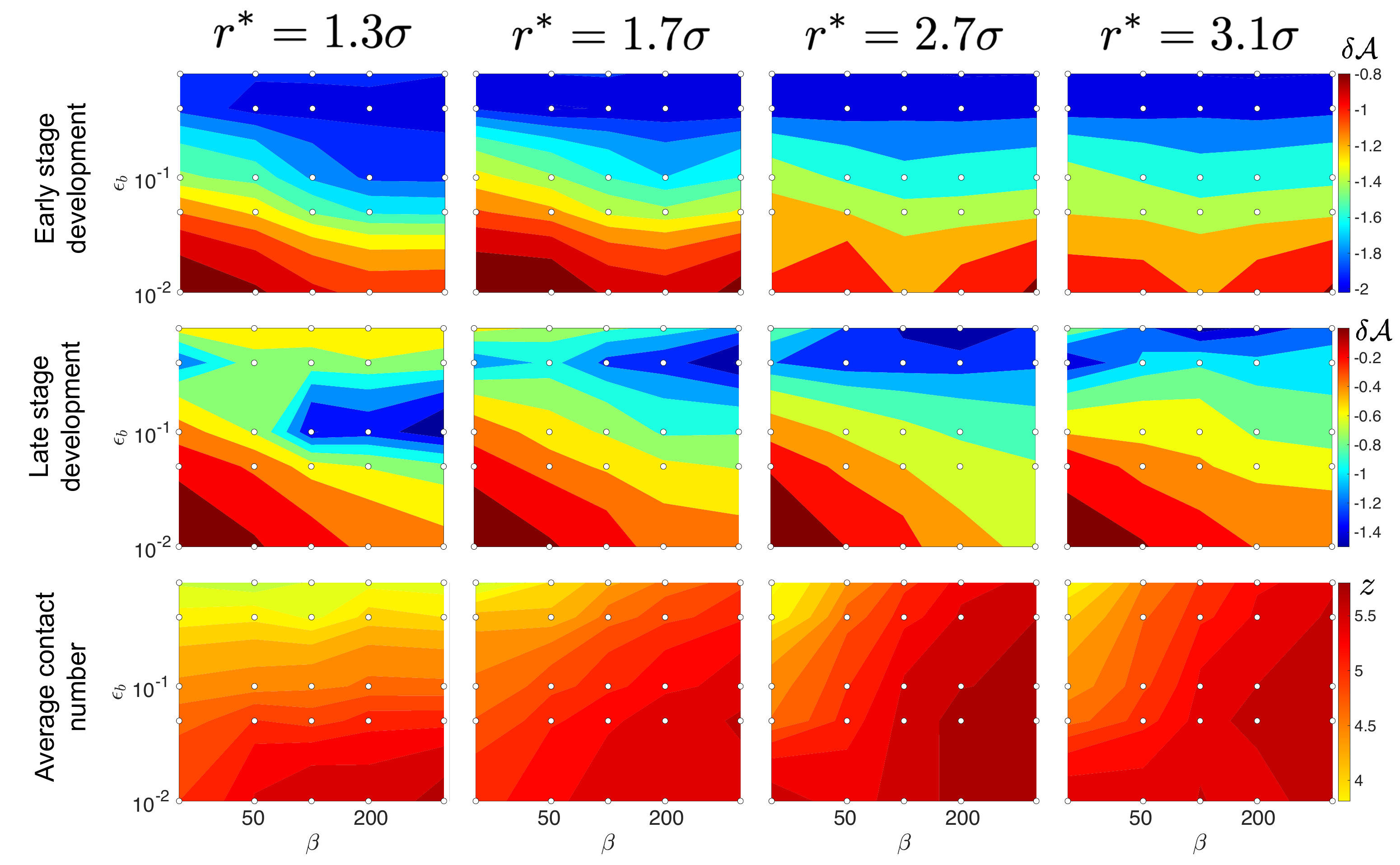}
    \caption{Effect of varying the bond breaking threshold $r^*$ from $1.3\sigma$ to $3.1\sigma$ on $\delta\mathcal{A}_{\rm early}$ (first row), $\delta\mathcal{A}_{\rm late}$ (second row), and $z$ at $\phi = 0.5$ (third row). The value of each color is given in the colorbar at the right of each row. Other parameters (defined in Table $1$ in the main text) for these simulations are $\Delta a=0.5$, $\lambda=5$, $c_L = 0.5$, $c_B = 4$, $\theta_{0,{\rm min}} = -3\pi/20$, and $P_0 = 10^{-6}$.}
    \label{appfig:hvar}
\end{figure}

In this section, we describe the details of the implementation of cell-cell adhesion in the DP model, as well as how adhesive bonds are formed and broken. Plant cells adhere to one another to perform a host of functions including those involved in tissue development~\citep{physio:SeagoAnnBot2005}. Adhesion between plant cells is maintained by a cellulose gel that lies between adherent neighbors and is actively cross-linked and regulated by a host of biochemical factors~\citep{physio:DaherFrontInPlantSci2015}. We do not attempt to recreate the complex viscoelastic properties of this gel, rather we represent adhesive forces between connected cells with harmonic bonds that break stochastically in response to stretching, and only near the edge of a junction between two cells. 

\subsection{Contact-dependent adhesion model}
As described in the main text, each developmental trajectory begins in a configuration of densely-packed deformable polygons, such that vertices on neighboring cells overlap. When two vertices on different cells overlap, say $i$ on cell $\mu$ and $j$ on cell $\nu$, we connect a harmonic bond between the two vertices. We track all vertices in the connectivity matrix $g$, where two vertices are bonded if $g_{ij\mu\nu} = 1$. 

We also assume that a given plant cell can only maintain a certain amount of adhesion at a given time, and that removing adhesive contacts strengthens remaining contacts by freeing resources like cellulose and regulatory proteins. That is, adhesion in our model is ``contact-dependent" and inversely proportional to the average number of contacts between two bonded cells. Contact-dependent adhesion is reasonable mechanism for the formation of porous, low-coordinated networks in mature tissues, as cells with fewer contacts will be able to "hold on" more tightly. For two vertices with contact distance $\sigma_{ij\mu\nu}$ and separation $r_{ij\mu\nu}$, we implement connectivity through the parameter $\epsilon_{ij\mu\nu}$ in Eq. $2$ in the main text, where
\begin{equation}\label{eq:epijmunu}
    \epsilon_{ij\mu\nu} = 
    \left\lbrace
        \begin{array}{ll}
            \epsilon_c &\text{ for }   r_{ij\mu\nu} < \sigma_{ij\mu\nu}\\
            z_{\mu\nu}^{-1} \epsilon_c  &\text{ for } r_{ij\mu\nu} > \sigma_{ij\mu\nu} \text{ and } g_{ij\mu\nu} = 1\\
            0 &\text{ for } r_{ij\mu\nu} > \sigma_{ij\mu\nu} \text{ and } g_{ij\mu\nu} = 0,
        \end{array}
    \right.
\end{equation}
where $\epsilon_c$ controls the strength of the interaction, $z_{\mu\nu} = (z_\mu + z_\nu)/2 + 1$, and $z_\mu$ is the number of cells in contact with cell $\mu$. Using Eq.~\eqref{eq:epijmunu} in Eq. $2$ in the main text amounts to a repulsive harmonic spring with stiffness $\sigma_{ij\mu\nu}^{-2}\epsilon_c$ when vertices overlap, but a bonded spring with stiffness $\sigma_{ij\mu\nu}^{-2}z_{\mu\nu}^{-1}\epsilon_c$ when two bonded vertices separate. 

\subsection{Bond-breaking}
During each enthalpy-minimization step, the bond connectivity $g$ is held fixed. Each bond is checked after enthalpy minimization and either maintained or broken. Bonds between two vertices may break if the energy required to maintain the bond exceeds a threshold. The potential energy stored in the bond between vertex $i$ on cell $\mu$ and $j$ on cell $\nu$ is 
\begin{equation}
    U_{ij\mu\nu} = \frac{\epsilon_c}{2z_{\mu\nu}}\qty(1 - \frac{r_{ij\mu\nu}}{\sigma_{ij\mu\nu}})^2.
\end{equation} 
A bond will then break if the energy exceeds a constant threshold $U_0$. Each bond will also have some probability to break spontaneously, either due to physical degradation of the bond by aging or active degradation of the bond by proteins~\citep{physio:LionettiPhytochem2015}. To model the effects of both spontaneous breaking and bond fracture, we define the probability $p_{\rm off}$ that a given bond breaks as
\begin{equation}\label{eq:bondbreak}
    p_{\rm off} = {\rm min}\qty[1,\exp(-\beta \Delta U)],
\end{equation}
where $\Delta U = U_0 - U_{ij\mu\nu}$, and $\beta$ controls bond robustness, i.e. if $\beta$ increases, bonds are more likely to stay intact. (See Fig. $2$D in the main text.) Our bond-breaking criterion is similar to the Metropolis algorithm~\citep{sim:AllenOxford2017}, where $\beta$ is similar to the inverse temperature. However, because bonds cannot reform and are only checked at successive enthalpy minima, our bond-breaking criterion does not bring the system to thermal equilibrium. 

We control the breaking threshold $U_0$ by setting the maximum distance, $r^*$, that a bond can be stretched before breaking. In our simulations, we set this distance to be $2.4\sigma$, which roughly corresponds to $U_0 = \epsilon_c/z_{\mu\nu}$ for contacts between cell $\mu$ and $\nu$. We show in Fig.~\ref{appfig:hvar} the effect of varying $r^*$ simultaneously with $\epsilon_b$ and $\beta$. We find that, while the simulation results for $\mathcal{A}$ and $z$ can change quantitatively as a function of $r^*$, our conclusions are unaffected if we were to choose a different value of $r^*$.

\section{Simulations of mesophyll tissue growth}\label{app:growth}
In this section, we describe the numerical methods we used to simulate the growth of spongy mesophyll tissue at constant boundary pressure. Our simulations are athermal and quasistatic, meaning that each small change in cell size is followed by relaxation to the nearest enthalpy minimum. We first describe the growth protocol that minimizes a given configuration's \emph{enthalpy}. Because pressure is conserved between each growth step, the simulation domain size fluctuates. We then detail how cell perimeters and areas are grown and curvatures are updated given local voids and intercellular contacts. 

\subsection{Athermal, quasistatic growth at constant pressure}
Our simulations minimize each configuration's enthalpy $H = U + P_0L_xL_y$, where $U$ is the total potential energy in Eq. $2$ in the main text, $P_0$ is the target pressure, and $L_x$ and $L_y$ are the side lengths of the simulation domain in the horizontal and vertical directions, respectively. The degrees of freedom for each configuration include the coordinates $\vec{r}_{i\mu}$ for the vertices $i=1,...,n_\mu$ on all cells $\mu = 1,...,N$. In this work, we only consider square domains that change size isotropically, i.e. $L_x = L_y = L$. However, this protocol can be extended to domains where $L_x$ and $L_y$ can vary independently. 

To minimize a given configuration's enthalpy $H$, we combine dynamics in the isoenthalpic-isobaric ensemble~\citep{constp} with enthalpy minimization using the FIRE algorithm~\citep{fire}. Isoenthalpic-isobaric dynamics (or NPH dynamics, as opposed to classical NVE dynamics with constant particle number, domain volume, and total internal energy) are obtained by treating the total domain size $A = L^2$ as an additional dynamical variable with mass $M$. In addition to momenta $\vec{p}_{i\mu}$ and masses $m_{i\mu}$ for all vertices $i$ on cells $\mu$, we define the domain ``momentum" $\Pi = M\pdv*{A}{t}$. The equations of motion for NPH dynamics in two spatial dimensions is:
\begin{equation}\label{eq:constHdynamics}
    \begin{split}
        \pdv{\vec{r}_{i\mu}}{t} &= \frac{\vec{p}_{i\mu}}{m_{i\mu}} + \frac{\Pi}{2MA}\vec{r}_{i\mu}\\
        \pdv{A}{t} &= \frac{\Pi}{M}\\
        \pdv{\vec{p}_{i\mu}}{t} &= -\pdv{U}{\vec{r}_{i\mu}} - \frac{\Pi}{2MA}\vec{p}_{i\mu}\\
        \pdv{\Pi}{t} &= P(t)-P_0,
    \end{split}
\end{equation}
where $P(t)$ is the instantaneous pressure at time $t$.  To compute the instantaneous pressure, we need to compute the total derivative of the potential energy $U$ (i.e. equations $1$ and $2$ in the main text) with respect to volumetric strains, i.e. to the domain boundary area $A$. To do this, first note that, for a square boundary with fixed box length $L$, the pressure can be written $P = -dU/dA = -(2L)^{-1}dU/dL$. The total derivative then becomes
\begin{equation}
    P = -\frac{1}{2L}\qty(\pdv{U}{L} + \sum_{i=1}^{N_v}\sum_{\xi = x,y}\pdv{U}{x_{i\xi}}\pdv{x_{i\xi}}{L})
\end{equation}
where we sum over all $N_v$ vertices in the system across all cells, and the $\xi$ indicates where $x_{i\xi}$ is either the $\xi = x$ or $\xi = y$ coordinate of vertex $i$. To measure the pressure, we perform a small deformation to the boundary where $L \to L + \Delta L$, and an affine transformation to the vertex coordinates $x_{i\xi} \to x_{i\xi}(L + \Delta L/L)$. In the limit $\Delta L \to 0$, we have $\pdv*{x_{i\xi}}{L} = x_{i\xi}/L$, and the pressure becomes
\begin{equation}\label{dpm:eq:P1}
    P = \frac{1}{2L^2}\sum_{i=1}^{N_v}\vec{F}_i \cdot \vec{r}_i - \frac{1}{2L}\pdv{U}{L}.
\end{equation}
The first term is referred to as the system's virial, which captures pressure in thermal systems or systems out of force balance~\citep{sim:AllenOxford2017}. As  The second term captures the purely potential contribution to the pressure, and needs to be derived more explicitly to be computed. We first note that
\begin{equation}\label{dpm:eq:dUdL1}
    \pdv{U}{L} = \sum_{\mu = 1}^N \qty[\pdv{U_{\rm a}}{a_\mu}\pdv{a_\mu}{L} + \sum_{i=1}^{n_\mu}\qty( \pdv{U_{\rm l}}{l_{i\mu}}\pdv{l_{i\mu}}{L} + \pdv{U_{\rm b}}{\theta_{i\mu}}\pdv{\theta_{i\mu}}{L} + \sum_{\nu\neq\mu}\sum_{j=1}^{n_\nu}\pdv{U_{\rm int}}{r_{ij\mu\nu}}\pdv{r_{ij\mu\nu}}{L})]. 
\end{equation}
The derivatives of potentials with respect to geometric parameters have already been found in order to compute forces. We then simply need to find how each geometric parameter changes with respect to deformation of the box size $L$. For lengths, i.e. $l_{i\mu}$ and $r_{ij\mu\nu}$, these objects behave identically to coordinates, therefore
\begin{equation}\label{dpm:eq:dldL}
    \pdv{l_{i\mu}}{L} = \frac{l_{i\mu}}{L} \qq{ , } \pdv{r_{ij\mu\nu}}{L} = \frac{r_{ij\mu\nu}}{L}.
\end{equation}
Because $\theta_{i\mu}$ is non-dimensional and does \emph{not} change with change of the simulation domain area, there is no bending energy contribution to the pressure.

To obtain the change in cell area with respect to box area, we note that the asphericity $\mathcal{A}$ of any given particle will also remain unchanged if the box area is uniformly grown, so therefore $\pdv*{\mathcal{A}}{L} = 0$. However, using the chain rule, we can show that a cell with area $a$ and perimeter $p$ must obey $\pdv*{a}{L} = (2a/p)\pdv*{p}{L}$. Because $\pdv*{p}{L} = p/L$ given that $p$ is composed of affinely deforming segment lengths, we must have that
\begin{equation}\label{dpm:eq:dadL}
    \pdv{a_\mu}{L} = \frac{2a_\mu}{L}.
\end{equation}
Using Eqs.~\eqref{dpm:eq:dldL} and~\eqref{dpm:eq:dadL} in Eq.~\eqref{dpm:eq:dUdL1}, we have
\begin{equation}
    \pdv{U}{L} = \sum_{\mu = 1}^N \left\lbrace \frac{2a_\mu}{a_{0\mu}L}\qty(\frac{a_\mu}{a_{0\mu}} - 1) + \sum_{i=1}^{n_\mu}\left[ \frac{\epsilon_l l_{i\mu}}{l_{0i\mu}L}\qty(\frac{l_{i\mu}}{l_{0i\mu}} - 1) - \sum_{\nu\neq\mu}\sum_{j=1}^{n_\nu}\frac{\epsilon_{ij\mu\nu}r_{ij\mu\nu}}{\sigma_{ij\mu\nu}L}\qty(1 - \frac{r_{ij\mu\nu}}{\sigma_{ij\mu\nu}})\right] \right\rbrace.
\end{equation}
Combining this expression with the virial term in Eq.~\eqref{dpm:eq:P1} gives a way to compute the instantaneous pressure in a configuration of plant cells given only the vertex coordinates, preferred geometric parameters and the box size.

While Eq.~\eqref{eq:constHdynamics} leads to dynamic sampling of the NPH ensemble, we require that each step in the simulations of spongy mesophyll development be at a local minimum of enthalpy. Standard implementation of the FIRE algorithm corresponds to potential energy minimization, where conservative forces $\vec{F}_{i\mu} = -\pdv*{U}{\vec{r}_{i\mu}}$ drive the system to minimize the potential energy, along with a force that increases momentum if the net force is decreasing in magnitude, but decreases momentum if the net force increases. The dynamical equations for systems in domains of fixed size are
\begin{equation}\label{eq:stdfire}
    \begin{split}
        \pdv{\vec{r}_{i\mu}}{t} &= \frac{\vec{p}_{i\mu}}{m_{i\mu}}\\
        \pdv{\vec{p}_{i\mu}}{t} &= \vec{F}_{i\mu} - \gamma(t)\qty(\vec{p}_{i\mu} - \frac{\pi}{\Phi}\vec{F}_{i\mu}),
    \end{split}
\end{equation}
where $\pi$ and $\Phi$ are the magnitudes of the vectors
\begin{equation}
    \begin{split}
        \vec{\pi} &= \mqty(p_{11,x} & p_{11,y} & p_{21,x} & \cdots & p_{n_N N,x} & p_{n_N N, y})\\
        \vec{\Phi} &= \mqty(F_{11,x} & F_{11,y} & F_{21,x} & \cdots & F_{n_N N,x} & F_{n_N N, y}),
    \end{split}
\end{equation}
i.e. the vectors of all momenta and net forces for all degrees of freedom, respectively. The variable $\gamma(t)$ is updated at each time step to tune whether the direction of $\vec{\pi}$ increases $\Phi$ (i.e. $\gamma>0$) or decreases $\Phi$ (i.e. $\gamma<0$).

To introduce dynamics that minimize the enthalpy, we include the domain size momentum $\Pi$ into the definition of $\vec{\pi}$, and the domain boundary force $P(t)-P_0$ into the definition of $\vec{\Phi}$. We also use the domain-adjusted forces $\vec{F}'_{i\mu} = -\pdv*{U}{\vec{r}_{i\mu}} - (\Pi/2MA)\vec{p}_{i\mu}$ in the global force vector $\vec{\Phi}$. The equations of motion for the FIRE algorithm to minimize enthalpy are
\begin{equation}\label{eq:NPHfire}
    \begin{split}
        \pdv{\vec{r}_{i\mu}}{t} &= \frac{\vec{p}_{i\mu}}{m_{i\mu}} + \frac{\Pi}{2MA}\vec{r}_{i\mu}\\
        \pdv{A}{t} &= \frac{\Pi}{M}\\
        \pdv{\vec{p}_{i\mu}}{t} &= \vec{F}'_{i\mu} - \gamma(t)\qty(\vec{p}_{i\mu} - \frac{\pi}{\Phi}\vec{F}'_{i\mu})\\
        \pdv{\Pi}{t} &= P(t)-P_0-\gamma(t)\qty[\Pi - \frac{\pi}{\Phi}\qty(P(t)-P_0)].
    \end{split}
\end{equation}
To solve these equations, we use an implementation and associated parameter values described in a recent assessment of FIRE~\citep{fire2} using a two-step velocity-Verlet algorithm~\citep{sim:AllenOxford2017}. We terminate the enthalpy-minimization process when $\vec{\Phi}$ reaches a sufficiently small magnitude, which corresponds to both small magnitudes for all net forces between degrees of freedom, as well as a small difference between $P(t)$ and $P_0$. As shown in Fig. $2$ in the main text, this protocol generates constant pressure at each step in the developmental simulations. For all simulations, we set the mass of each vertex and the domain boundary mass to be equal to $1$. 

\subsection{Updates to area, perimeter, and curvature}
After each enthalpy-minimization step, we simulate growth of cell areas and void-facing perimeters, and update the curvature of void-facing vertices,
\begin{equation}
    \begin{split}
        a_0 &\to a_0 + \Delta a\\
        l_0 &\to l_0 + \Delta l\\
        \theta_0 &\to \theta_0 - \Delta \theta.
    \end{split}
\end{equation}
As mentioned in the main text, areal growth $\Delta a$ is a constant. We define the basal rate of growth of the void-facing perimeter segments $\Delta \tilde{l}$ as
\begin{equation}\label{eq:dldef}
    \Delta \tilde{l}(\lambda) = \frac{\lambda p\Delta a}{2na},
\end{equation}
where $\lambda$ is a dimensionless quantity that sets the rate of growth, $n$ is the number of vertices on a given DP model cell, $a$ is the instantaneous cell area, and $p$ is the instantaneous cell perimeter. This scaling is chosen so that, if $\lambda = 1$ and all perimeter segments are void-facing (i.e. if a given cell had no contacts), each cell grows isometrically and preserves its shape. In addition to Eq.~\eqref{eq:dldef}, we include an adjustment to the basal rate of growth such that no individual void-facing segment grows too fast compared to the other segments on a given cell. If a given void-facing segment participates in a total void-facing boundary of length $b_v$ (made up of at least one or more adjacent void-facing segments, see e.g. the red segments in Fig. $2$F in the main text), we grow that segment according to the rate
\begin{equation}
    \Delta l(\lambda,c_L) = \Delta \tilde{l}(\lambda)\qty[1 + c_L\qty(\frac{\bar{b}_v - b_v}{n} - 1)],
\end{equation}
where $\bar{b}_v$ is the average length of void-facing boundaries on a given cell. We choose $0 \leq c_L \leq 1$ such that $c_L = 0$ corresponds to pure growth, and $c_L = 1$ corresponds to relaxational dynamics towards the mean void boundary length $\bar{b}_v$. Our choice for this form of the perimeter growth rate assumes that cells actively track their mean void boundary length during development, and can adjust the rate of growth of void-facing surfaces accordingly. Given that microtubles are known to accrue at void-facing boundaries in spongy mesophyll cells~\citep{physio:ZhangPlantCell2021}, there may exist ways in which cells might measure this quantity, but to our knowledge no specific mechanism is known. We set $c_L = 0.5$ throughout this study to balance cell shape regularization and growth. We have independently verified that small changes to this parameter do not significantly affect the results presented here, but have found that values close to both $0$ and $1$ do not lead to robust porous network formation. 

In addition to growing cell areas and void-facing segments, we also tune cell shapes to take on more lobed-like structures by driving the spontaneous curvature $\theta_0$ toward a negative, constant value $\theta_{0,{\rm min}}$ using angular increment $\Delta\theta$. When choosing $\Delta\theta$, we assume that the rate of curvature change is proportional to the rate of growth. We therefore define the change in the intrinsic bending angle at vertex $k$ on cell $\mu$ as
\begin{equation}~\label{eq:dtheta}
    \Delta\theta_{k\mu}(c_B) = \frac{\Delta \bar{l}_{k\mu}(1)}{\sqrt{a_{0\mu}}}\qty[c_B - \qty(\theta_{k+1,\mu}-\theta_{k\mu}) - \qty(\theta_{k-1,\mu}-\theta_{k\mu})],
\end{equation}
where $\Delta\bar{l}(1)$ is defined in Eq.~\eqref{eq:dldef} and $c_B$ is a free parameter that sets the magnitude of the growth rate. The last two terms in Eq.~\ref{eq:dtheta} relax the value of $\theta_k$ toward the values of $\theta_{k-1}$ and $\theta_{k+1}$, which tends to smooth cell boundary curvature. In this work, we set $c_B = 4$, which drives curvature toward negative values at an optimal rate for porous network formation. We have independently varied both the rate of curvature smoothing during growth, as well as the values of $c_B$ and $\theta_{0,{\rm min}}$, and verified that our results do not depend sensitively on small changes to the values of the parameters we used in the simulations. However, if we set $\theta_{0,{\rm min}} > 0$ and $c_B \approx 1$, these parameter selections do not yield robust porous networks.

\bibliography{mesosim}